\documentclass[11pt, dina4]{article}
\usepackage{amssymb, amsmath,amsthm}
\usepackage{graphicx}
\usepackage{multirow}
\usepackage{cite}
\usepackage{multirow}
\usepackage{epstopdf}
\usepackage{float}
\usepackage{subfig}
\usepackage{color}
\usepackage[margin=1in]{geometry}
\usepackage{textcomp}
\usepackage{amsmath}

\newcommand{\p}{\partial}

\newcommand{\Og}{\Omega}
\newcommand{\fl}[2]{\frac{#1}{#2}}

\newcommand{\gm}{\gamma}

\newcommand{\Dt}{\Delta}

\newcommand{\be}{\begin{equation}}
\newcommand{\ee}{\end{equation}}
\newcommand{\ba}{\begin{array}}
\newcommand{\ea}{\end{array}}
\newcommand{\bea}{\begin{eqnarray}}
\newcommand{\eea}{\end{eqnarray}}
\newcommand{\beas}{\begin{eqnarray*}}
\newcommand{\eeas}{\end{eqnarray*}}

\newcommand{\bx}{{\bf x} }

\newcommand{\bb}{\bigskip}
\newcommand{\bu}{{\bf u}}
\newcommand{\bk}{{\bf k}}
\newcommand{\bveps}{\mbox{\boldmath$\varepsilon$}}
\newcommand{\veps}{\varepsilon}

\title{Pattern selection in the Schnakenberg equations: From normal to anomalous diffusion}
\author{
Hatim K. Khudhair\thanks{Department of Mathematics and Statistics, Missouri University of Science and Technology, Rolla, MO 65409-0020 (Email: hkkz89@mst.edu) },  \ \ 
Yanzhi Zhang\thanks{Department of Mathematics and Statistics, Missouri University of Science and Technology, Rolla, MO 65409-0020 (Email: zhangyanz@mst.edu)}, \ \ 
Nobuyuki Fukawa\thanks{Department of Business and Information Technology, Missouri University of Science and Technology, Rolla, MO 65409-0020 (Email: fukawan@mst.edu)}}
\date{}

\begin{document}
\maketitle

\begin{abstract}
Pattern formation in the classical and fractional Schnakenberg equations is studied to understand the nonlocal effects of anomalous diffusion. 
Starting with linear stability analysis,  we find that if the activator and inhibitor have the same diffusion power, the Turing instability space depends only on the ratio of diffusion coefficients $\kappa_1/\kappa_2$. 
However, the smaller diffusive powers might introduce larger unstable wave numbers with wider band,  implying that the patterns may be more chaotic in the fractional cases. 
We then apply a weakly nonlinear analysis to predict the parameter regimes for spot,  stripe, and mixed patterns in the Turing space. 
Our numerical simulations confirm the analytical results and demonstrate the differences of normal and anomalous diffusion on pattern formation.
We find that in the presence of superdiffusion the patterns exhibit multiscale structures. 
The smaller the diffusion powers, the larger the unstable wave numbers and the smaller the pattern scales. 
 \end{abstract}


\section{Introduction}
\label{section1}
\setcounter{equation}{0}

The reaction-diffusion equations have wide applications in many fields, including biology, chemistry, ecology, geology, physics, finance, and so on. 
In classical reaction-diffusion equations, the diffusion is described by the standard Laplace operator $\Dt = \p_{xx} + \p_{yy} + \p_{zz}$, characterizing the transport mechanics due to the Brownian motion. 
Recently, it has been suggested that many complex (e.g., biological and chemical) systems are indeed characterized by the L\'evy motion, rather than the Brownian motion; see \cite{Metzler2000, Cusimano2015, del--Castillo--Negrete2003, Giona1992, Hofling2013, Hornung2005, Naether2013, Ren2003, Shlesinger1987, Zimbardo2015} and references therein. 
Hence, the fractional reaction-diffusion equations were proposed to model these systems, where the  L\'evy anomalous diffusion is described by the fractional Laplacian $(-\Dt)^\gm$. 
So far, many studies can be found on the fractional reaction-diffusion equations \cite{BuenoOrovio2014, Gafiychuk2008, Gafiychuk2010, Golovin2008}. 
However, the anomalous diffusion and its interplay with nonlinear reactions on pattern formation and selection have not been well understood yet. 

In this paper, we study the pattern formation and selection in the Schnakenberg equation to compare  the effects of normal and anomalous diffusion.
The Schnakenberg equation is  one of the simplest reaction-diffusion systems. 
It has been applied to study pattern formation in, such as animal skins \cite{Shaw1990, Barrio2009, Aragon1998},  plant root hair initiation \cite{Li2018},  and fluid flows \cite{Wei2012}. 
The Schnakenberg equation was first introduced in \cite{Schnakenberg1979} to study the limit cycle behavior, i.e., temporal periodic solutions. 
It describes the following reactions:  $ A\rightleftharpoons X$,  $B\to Y$ and $2X+Y \to3X$, with $X$ $Y$, $A$ and $B$ representing different chemicals. 
In this study, we consider the  Schnakenberg equation of the following form:  
\bea\label{SNG}
\begin{array}{l}
\displaystyle\p_t u(\bx, t) = -\kappa_1(-\Dt)^{\fl{\gamma_1}{2}}u  + A-u + u^2v, \\
\displaystyle \p_t v(\bx, t) = -\kappa_2 (-\Dt)^{\fl{\gamma_2}{2}}v + B- u^2v, 
\end{array}
\eea
where $u$ and $v$ denote the concentration of chemicals $X$ and $Y$, respectively, and  $\kappa_1,\kappa_2$ are their diffusion coefficients.
With a slight abuse of notation, we denote $A$ and $B$ as the concentrations of chemical $A$ and $B$, respectively.  
We assume that $A$ and $B$ are in abundance so that $A$ and $B$ are kept constant in the model (\ref{SNG}). 
The fractional Laplacian $(-\Dt)^{\fl{\gm}{2}}$ is defined as 
\beas
(-\Dt)^{\fl{\gm}{2}}u = {\mathcal{F}}^{-1}\big[|{\bf k}|^{\gamma} {\mathcal{F}}\big[u\big]\big], \qquad \mbox{for} \ \ \gamma  > 0,
\eeas
where ${\mathcal{F}}$ represents the Fourier transform, and ${\mathcal{F}}^{-1}$ denotes the associated inverse transform.  
Probabilistically, the fractional Laplacian represents the infinitesimal generator of a symmetric $\gm$-stable L\'evy process. 
In the special case of $\gm_1 = \gm_2 = 2$, the system (\ref{SNG}) reduces to the classical Schnakenberg equation \cite{Schnakenberg1979}.
In this study, we are interested in the diffusion power $ \gamma_1, \gamma_2  \in (1, 2]$. 

Pattern formation and pattern selection have  been one of the most important topics in the study of  reaction-diffusion equations. 
For the classical Schnakenberg equations, many theoretical results have been reported in the literature, including the existence of steady states \cite{Li2011, Li2017, Li2018}, 
various Turing patterns and their stability  \cite{Ward2002, Beentjes2014, Iron2004, Liu2013, Liu2017, Kolokolnikov2009, Gomez2020}, and Hopf bifurcation analysis \cite{Xu2012, Yi2017}. 
In contrast, the study of the fractional Schnakenberg equation still remains scant. 
In \cite{Hammouch2014}, a finite difference method is proposed to solve the variable-order fractional Schnakenberg equations. 
Recently, there is growing interest in the fractional reaction-diffusion equations (see \cite{BuenoOrovio2014,  Dutt2012, Golovin2008, Prytula2016, Gafiychuk2008, Zhang2014} and reference therein), 
but the understanding of anomalous diffusion in pattern formation and selection still remains limited.
To the best of our knowledge, no report on pattern formations in the fractional Schnakenberg equation can be found in the literature. 
Moreover, even though there are many theoretic studies on the classical Schnakenberg equation, few numerical studies can be found on pattern formations. 

In this work, we analytically and numerically study pattern formation and selection in both classical and fractional Schnakenberg equations. 
As one of the simplest reaction-diffusion systems, the study of pattern formations in the Schnakenberg equation provides insights to understand anomalous diffusion in reaction-diffusion models and advances their practical applications.
We find that the necessary condition for Turing instabilities is $\kappa_1 < \kappa_2$.
If $\gamma_1 = \gamma_2$,  the fractional Schnakenberg equations have the same Turing spaces as its classical counterpart, but the Turing space increases with the ratio $\kappa_1/\kappa_2$ or $\gamma_1/\gamma_2$. 
The smaller the power  $\gamma_1$, the larger the unstable wave numbers and the smaller the pattern scales.  
Our weakly nonlinear analysis predicts the parameter regimes for hexagon patterns, stripe patterns, and their mixtures.  
Our numerical results confirmed the theoretical analysis and also provided new insights on the patterns in the fractional Schnakenberg equations. 
This paper is organized as follows. 
In Sect. \ref{section2}, we carry out a linear stability analysis to study Turing instability. 
In Sect. \ref{section3},  weakly nonlinear stability analysis and amplitude equation analysis are presented to study the patterns of hexagons, stripes, and their coexistence. 
Numerical studies of patterns in the classical Schnakenberg equation are presented in Sect. \ref{section4}, while Sect. \ref{section5} is devoted to patterns in the fractional cases. 
Finally, some conclusions are made in Section \ref{section6}.

\section{Linear stability analysis}
\label{section2}
\setcounter{equation}{0}

In this section, we perform the linear stability analysis for the Schnakenberg model (\ref{SNG}) and study the conditions for Hopf and Turing bifurcations. 
For notational convenience,  we let ${\bf u} = (u, v)^T$ and  denote $a = B-A$ and $b = B+A$;  thus the system (\ref{SNG}) can be reformulated as: 
\bea\label{SNG1}
\begin{array}{l}
\displaystyle\p_t u(\bx, t) = -\kappa_1(-\Dt)^{\fl{\gamma_1}{2}}u-u + u^2v+\fl{b-a}{2}, \\
\displaystyle \p_t v(\bx, t) = -\kappa_2 (-\Dt)^{\fl{\gamma_2}{2}}v -u^2v + \fl{b+a}{2}. 
\end{array}
\eea
Noticing that $A, B > 0$, we have $b > 0$ and $a \in (-b, b)$ in (\ref{SNG1}).  
In the absence of diffusion (i.e., $\kappa_1=\kappa_2=0$),  the system (\ref{SNG1}) has a unique  stationary state ${\bf u}_s \equiv \big(b, \ (a+b)/2b^2\big)^T$.  
Furthermore,  the Jacobian matrix of system (\ref{SNG1}) at $\bu_s$ is given by
\beas 
J|_{{\bf u} = {\bf u}_s}=\left(\begin{array}{ccc} 
a/b  && b^2\\
-\big(1+a/b\big) && -b^2\end{array}\right). 
\eeas
It is evident that the  steady state $\bu_s$ is stable,  if the   trace and determinant of $J$ satisfy
\beas
{\rm tr}(J) = \fl{a}{b} - b^2 < 0, \qquad {\rm det}(J) = b^2 > 0,
\eeas
equivalently, we require $a < b^3$. 

Next, we carry out linear stability analysis to understand the stability of $\bu_s$ in the presence of diffusion (i.e., $\kappa_1, \kappa_2 \neq 0$). 
Consider a small perturbation of the steady state $\bu_s$, i.e., 
\bea\label{pertub}
{\bf u} = {\bf u}_s  \, + \mbox{\boldmath$\varepsilon$}\exp(\lambda t + i {\bf k}\cdot \bx),
\eea
where $\bveps = (\veps_1, \veps_2)^T$ with $|\veps_1|, \, |\veps_2| \ll 1$ being the amplitudes of perturbations,  $i = \sqrt{-1}$ is the imaginary unit, $\lambda$ is the growth rate of the perturbation in time $t$, and ${\bf k}$ is the  wave vector.  
Substituting (\ref{pertub}) into (\ref{SNG1}) and linearizing the system, we  obtain
\beas
\left(\begin{array}{ccc} 
\lambda + \kappa_1|{\bf k}|^{\gm_1} - a/b & & -b^2 \\
1 + a/b & & \lambda + \kappa_2|{\bf k}|^{\gm_2} + b^2
\end{array}\right)
\left(\begin{array}{c} 
\veps_1 \\ \veps_2
\end{array}\right) 
= \left(\begin{array}{c} 
0\\0
\end{array}\right).
\eeas
Hence, the characteristic equation of the above system is
\bea\label{char}
\lambda^2 - \left(\frac{a}{b}-b^2 -\kappa_1|{\bf k}|^{\gamma_1} - \kappa_2|{\bf k}|^{\gamma_2}\right)\lambda + \Big[\Big(\kappa_1|{\bf k}|^{\gamma_1} - \fl{a}{b}\Big)\Big(\kappa_2|{\bf k}|^{\gamma_2} + b^2\Big) + b^2\Big(1+\fl{a}{b}\Big)\Big] = 0. \eea
The Hopf bifurcation occurs when $|\bf k| = 0$ and ${\rm Re}(\lambda) = 0$,  but ${\rm Im}(\lambda)\neq 0$. 
Thus, the boundary of Hopf bifurcation is given by 
\beas
a = b^3.
\eeas

If ${\rm Re}(\lambda({\bf k})) > 0$, the unstable wave number ${\bf k}$ will grow exponentially until the nonlinearity bounds this growth. 
The onset of the instability  occurs at $\lambda({\bf k}) = 0$, i.e., when 
\beas
\Big(\kappa_1|{\bf k}|^{\gamma_1} - \fl{a}{b}\Big)\Big(\kappa_2|{\bf k}|^{\gamma_2} + b^2\Big) + b^2\Big(1+\fl{a}{b}\Big) = 0,
\eeas
which admits the single minimum $(k_{\rm cr}, \, a_{\rm cr})$:
\bea\label{critical}
 a_{\rm cr} = \fl{\kappa_1\sqrt{\kappa_1\kappa_2\sigma}\big(\kappa_1\vartheta + \sigma + 1\big)}{\big[\kappa_1 \vartheta (1-\sigma) +1\big]^{3/2}}\,\vartheta^{\fl{3}{2}+\fl{1}{2\sigma}}, \qquad 
k_{\rm cr} = |{\bf k}_{\rm cr}| = \vartheta^\fl{1}{\gm_1}.
\eea
Here, we denote  $\sigma = \gamma_1/\gamma_2$, implying that $\sigma \in (\fl{1}{2},  2)$  as $\gamma_1, \gamma_2 \in (1, 2]$.
We define $\vartheta$ implicitly as:
\beas
\fl{\kappa_1 \kappa_2\,\sigma\,\vartheta^{1+1/\sigma}}{\kappa_1(1-\sigma)\vartheta + 1} = b^2, 
\eeas
which implies that $\vartheta > 0$ for $\sigma \in (\fl{1}{2}, 1]$, or $0 < \vartheta < 1/\kappa_1(\sigma-1)$ for $\sigma \in (1, 2)$. 
In the special case of $\gamma : =\gamma_1 = \gamma_2$, we have  $\sigma = 1$ and $\vartheta = b/\sqrt{\kappa_1\kappa_2}$, and thus the critical values in (\ref{critical}) reduce to 
\bea\label{critical0}
a_{\rm cr}=b^2 r (br+2), \qquad k_{\rm cr} = \bigg(\frac{b}{\sqrt{\kappa_1\kappa_2}}\bigg)^{1/\gm},
\eea
where we denote $r = \sqrt{\kappa_1/\kappa_2}$, i.e., the square root of the diffusion coefficient ratio.

From the above discussion and noticing $a \in (-b, b)$, we see that the conditions for  the Turing instability (also known as diffusion-driven instability) are given by
\bea \label{turing}
a_{\rm cr} < a < \min\{b,  b^3\},
\eea
which is referred to as the Turing space. 
The conditions (\ref{critical})--(\ref{turing}) suggest that the Turing space generally depends on the diffusion coefficients $\kappa_1$ and $\kappa_2$,  and the ratio $\sigma$ of diffusion powers. 
Particularly, if $\gamma_1 = \gamma_2$ this dependence reduces to the ratio of diffusion coefficients (i.e., $\kappa_1/\kappa_2$), rather than the values of $\kappa_1$ and $\kappa_2$. 
Comparing (\ref{critical0}) and (\ref{turing}) suggests that for $\gamma_1 = \gamma_2$, the necessary condition of Turing instability is $\kappa_1 < \kappa_2$, i.e., the inhibitor $v$ should diffuse faster. 

\begin{figure}[htb!]
\centerline{
(a)\includegraphics[height=5.560cm,width=7.860cm]{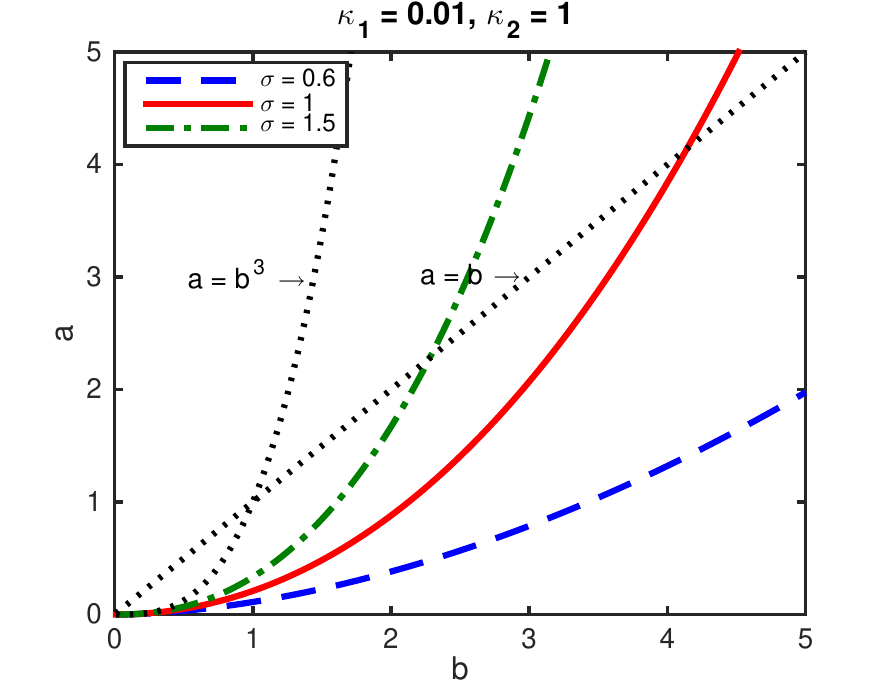}\hspace{-5mm}
(b)\includegraphics[height=5.560cm,width=7.860cm]{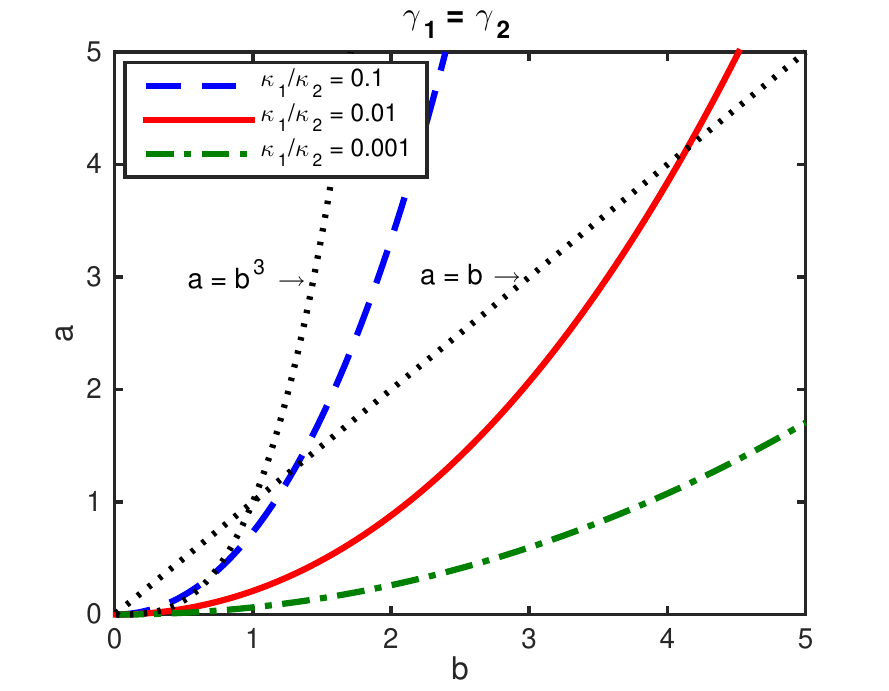}}
\caption{Illustration of Turing spaces (i.e., region between the non-dotted line and dotted-line of $\min\{b,\, b^3\}$) for different ratios of diffusion powers  $\sigma = \gm_1/\gm_2$ (a) and coefficients $\kappa_1/\kappa_2$ (b). }
\label{Fig1}
\end{figure}

Figure \ref{Fig1} illustrates the Turing spaces for different parameters.  
In Fig. \ref{Fig1}(a), we fix the diffusion coefficients $\kappa_1$ and $\kappa_2$, and compare the Turing spaces for different ratios $\sigma$ of diffusive powers.
It shows that with the increase of $\sigma$, the Turing space reduces quickly. 
Fig. \ref{Fig1}(b) shows the effects of diffusion coefficients for  $\gm_1 = \gm_2$. 
We find  that the smaller the ratio $\kappa_1/\kappa_2$, the larger the Turing space, 
and   our extensive studies show that this observation is independent of $\sigma$.
In the special case of $\gamma_1 = \gamma_2$, to ensure the existence of Turing space, the diffusion ratio $r$ should satisfy $r \le \big(-1 + \min\big\{\sqrt{2},  \sqrt{1+b^{2}} \big\}\big)/b$, which implies that the maximum ratio allowed depends on parameter $b$.

\begin{figure}[htb!]
\centerline{
(a)\includegraphics[height=5.260cm,width=7.460cm]{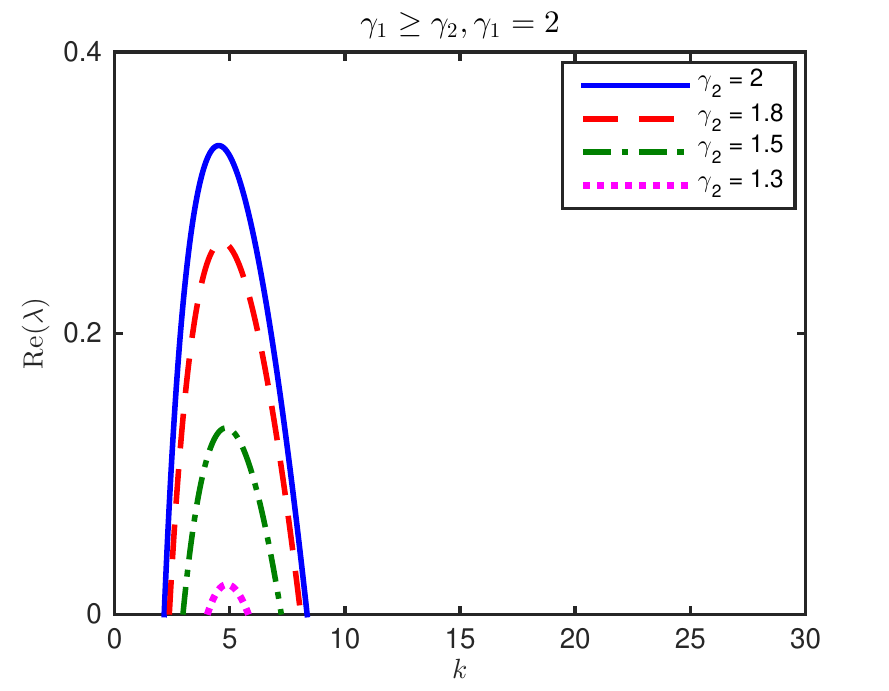}\hspace{-3mm}
(b)\includegraphics[height=5.260cm,width=7.460cm]{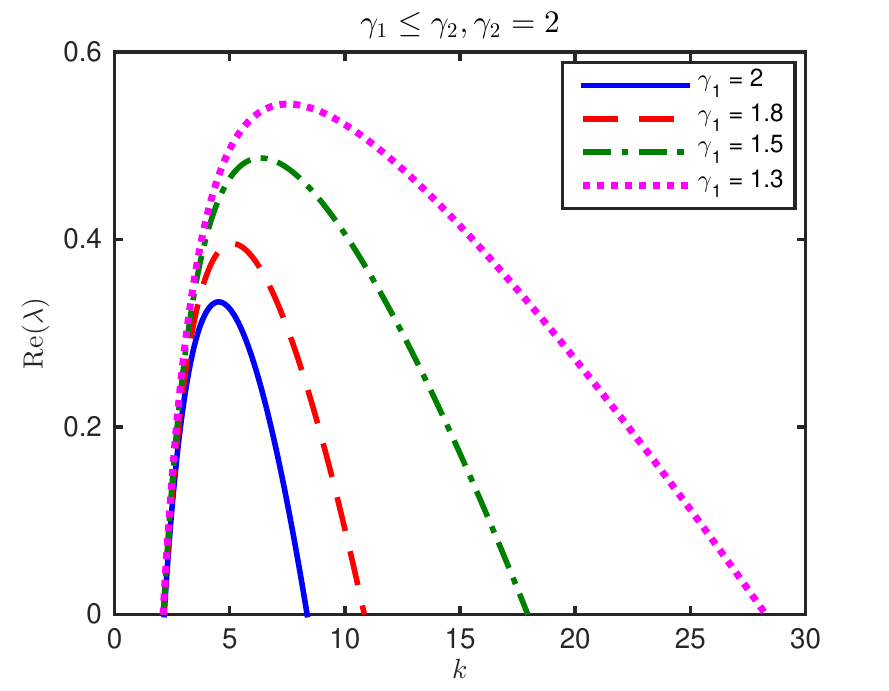}}
\centerline{
(c)\includegraphics[height=5.260cm,width=7.460cm]{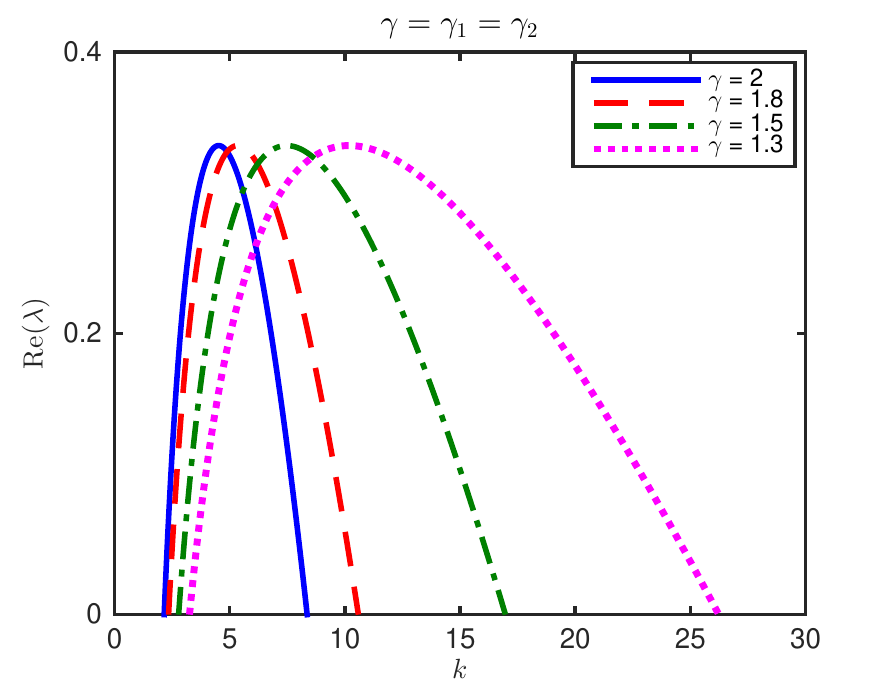}\hspace{-3mm}
(d)\includegraphics[height=5.260cm,width=7.460cm]{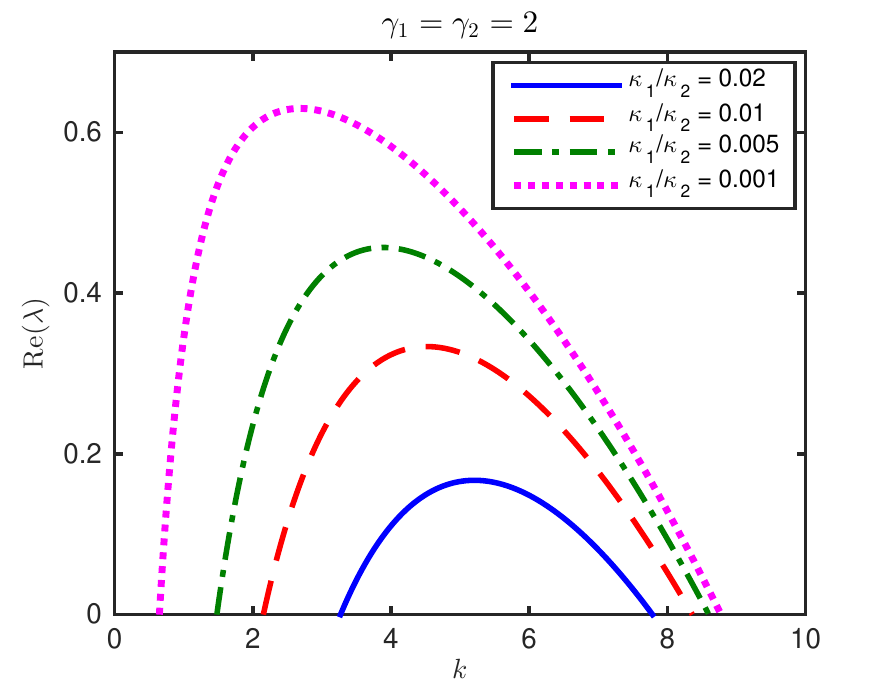}}
\caption{Illustration of growth rates for different parameters, where $a = 1.4$ and $b = 1.8$ are fixed.  
The diffusion coefficients $\kappa_1 = 0.01$ and $\kappa_2 = 1$ for (a) $\gamma_1 \ge \gamma_2$;  (b) $\gamma_1 \le \gamma_2$; (c) $\gamma_1 = \gamma_2$. }
\label{Fig2}
\end{figure}

Figure \ref{Fig2} compares the unstable bands of wave number $k = |{\bf k}|$ for different diffusive parameters. 
Fig. \ref{Fig2}(a)--(c)  show that for given $\kappa_1$ and $\kappa_2$, the ratio $\sigma$ of diffusion powers play an important role in determining the maximum growth rate (i.e., $\max_k{\rm Re}(\lambda)$) of the unstable wave number. 
The maximum growth rate decreases with increasing ratio $\sigma = \gm_1/\gm_2$, and thus maximum growth rates for both classical and fractional cases remain the same as long as $\gamma_1 = \gamma_2$; see Fig. \ref{Fig2}(c). 
It further suggests that the Turing instability could occur even if $\gamma_1 \ge \gamma_2$, but not for $\kappa_1 \ge \kappa_2$.
On the other hand, the width of unstable bands depends on the powers $\gamma_1$ and $\gamma_2$, rather than their ratio. 
Fig. \ref{Fig2} (c) shows that if $\gamma_1 = \gamma_2$, the fractional Schnakenberg equation has more unstable wave numbers than its classical counterpart. 
Moreover, the instability tends to occur at larger wave numbers. 
 Fig. \ref{Fig2} (d) additionally compares the unstable wave numbers for different diffusion ratio $\kappa_1/\kappa_2$, where $\gamma_1 = \gamma_2 = 2$.
It shows that the decrease of diffusion ratio $\kappa_1/\kappa_2$ broadens the unstable band and also increases the maximum growth rate. 
Even though decreasing the ratio $\kappa_1/\kappa_2$ or the fractional power $\gamma$ both lead to a wider unstable band, they are essentially different diffusion mechanics (cf. Fig. \ref{Fig2} (c) \& (d)).

\section{Weakly nonlinear analysis}
\label{section3}
\setcounter{equation}{0}

The linear stability analysis predicts unstable wave numbers in the system, but it fails to provide insights on the nonlinear coupling of these unstable wave numbers. 
In the study of pattern formation, however, the nonlinear terms dominate the growth of the unstable modes. 
In this section, we will perform a weakly nonlinear analysis of the system (\ref{SNG1}) near the Turing instability threshold $(k_{\rm cr}, \, a_{\rm cr})$, where the solution of (\ref{SNG1}) can be written in the form 
\bea\label{solution0623}
\bu = \bu_s +  \sum_{1 \le j \le 3} \left({\bf A}_j(t)\,e^{i {\bf k}_j \cdot \bx} + \bar{\bf A}_j(t)\,e^{-i {\bf k}_j \cdot \bx}\right)
\eea
where ${\bf A}_j = \big({A}_{j,u}, A_{j,v}\big)^T$ denotes the amplitude associated with wave number ${\bf k}_j$, and $\bar{\bf A}_j$ represents its complex conjugate. 
The wave number satisfies ${\bk}_j = k_{\rm cr} (\cos(2j\pi/3), \, \sin(2j\pi/3))^T$ (for $1 \le j \le 3$), and thus $\bk_1 + \bk_2 + \bk_3 = {\bf 0}$.
In the following, we will denote $e_j = e^{i {\bf k}_j \cdot \bx}$ for notational simplicity, and will focus on the analysis of $\gm_1 = \gm_2 = \gm$. 

Next, we derive the amplitude equations for ${\bf A}_j$. 
Introducing the slow time  $\tau = \veps^2t$,  we then expand $\bu - \bu_s$ and the bifurcation parameter $a$ as:
\bea\label{expand-u}
\bu - \bu_s = \veps \bu_1 + \veps^2 \bu_2 + \veps^3 \bu_3 + {\mathcal O}(\veps^4), \qquad \ 
a = a_{\rm cr} + \veps^2 \hat{a} + {\mathcal O}(\veps^3),
\eea
where $\bu_i := \bu_i(\bx, \tau) = (u_i, v_i)^T$ for $i = 1, 2, 3$. 
Substituting (\ref{expand-u}) into (\ref{SNG1}), and collecting like powers of $\veps$, we obtain the sequence of equations as
\bea
\label{veps1}
{\mathcal O}(\veps):  && {\mathcal L}\bu_1 = {\bf 0}, \\
\label{veps2}
{\mathcal O}(\veps^2):  && {\mathcal L}\bu_2 =  {\bf c}\bigg(2bu_1 v_1 + \fl{1}{2b}\left(1+\fl{a_{\rm cr}}{b}\right)u_1^2\bigg)
, \\
\label{veps3}
{\mathcal O}(\veps^3):  && {\mathcal L}\bu_3 = \p_\tau {\bf u}_1 + {\bf c}\bigg(u_1^2v_1 + 2b (u_1v_2+u_2 v_1) + \fl{1}{b}\left(1+\fl{a_{\rm cr}}{b}\right)u_1 u_2 + \fl{\hat{a}}{b}u_1\bigg),\qquad\quad 
\eea
where  the vector ${\bf c} = (-1, 1)^T$,  and ${\mathcal L}$ denotes the linear operator of the system at the Turing instability threshold, i.e.,
\beas\label{mathcal L}
{\mathcal L} = \left(\begin{array}{ccc}
\displaystyle -\kappa_1(-\Dt)^{\fl{\gamma}{2}} +\frac{a_{\rm cr}}{b} && b^2 \\
\displaystyle-\Big(1+\frac{a_{\rm cr}}{b}\Big)&& -\kappa_2(-\Dt)^{\fl{\gamma}{2}} -b^2 
\end{array}\right).
\eeas

At ${\mathcal O}(\veps)$, we seek the solution of (\ref{veps1}) of the form: 
\bea\label{u1}
\bu_1 = \begin{pmatrix} \hat{u}_1 \\ \hat{v}_1 \end{pmatrix} \sum_{1 \le j \le 3} \Big(W_j(\tau)\,e_j + \bar{W}_j(\tau)\,\bar{e}_j \Big),
\eea
where $W_j$ denotes the amplitude of the wave number ${\bf k}_j$ at the first order perturbation (i.e., at ${\mathcal O}(\veps)$).
Substituting (\ref{u1}) into (\ref{veps1}) and noticing the values of $a_{\rm cr}$ and $k_{\rm cr}$ in (\ref{critical0}), we obtain 
\bea\label{rho}
\hat{u}_1 = b, \qquad \hat{v}_1 = -r(br + 1). 
\eea

At ${\mathcal O}(\veps^2)$, we can rewrite (\ref{veps2}) as: 
\bea\label{veps2-1}
{\mathcal L}\bu_2 = {\bf c}\,\xi \left(\Theta + \bar{\Theta}\right)^2  = {\bf c}\,\xi\sum_{1 \le j \le 4}\big(\Theta_j + \bar{\Theta}_j\big),
\eea
by taking (\ref{u1}) and (\ref{rho}) into account, 
where we denote
\beas
&&\xi =  \fl{b}{2} - b^2 r \Big(1+ \fl{3}{2}b r\Big), \quad \ \ \Theta = \sum_{1 \le j \le 3} W_j e_j, \quad\ \   
 \Theta_1 = |{\bf W}|^2 := \sum_{1 \le j \le 3} |W_j|^2, \\
&&\Theta_2 =  \sum_{1 \le j \le 3} W_j^2\,e_j^2, \quad \ \  \Theta_3 = 2 \sum_{\substack{j = 1, 2 \\ j < l \le 3} }W_j \bar{W}_l \, e_j\bar{e}_l, \quad \ \ \Theta_4 = 2 \big(W_1 W_2\,\bar{e}_3+W_1 W_3\,\bar{e}_2 +W_2 W_3\,\bar{e}_1\big).  \qquad 
\eeas
The $\Theta_4$ terms are introduced by the resonant interactions between models $e_j$, which are usually considered  to be small \cite{Golovin2008}.  
Hence, we will neglect the terms of $\Theta_4$ in the following discussion.  
Solving (\ref{veps2-1}) gives the solution $\bu_2$ as 
\bea\label{u2}
\bu_2 = \xi\bigg[\begin{pmatrix} \hat{u}_{21} \\ \hat{v}_{21} \end{pmatrix} \big(\Theta_1 + \bar{\Theta}_1\big)+\begin{pmatrix} \hat{u}_{22} \\ \hat{v}_{22} \end{pmatrix} \big(\Theta_2 + \bar{\Theta}_2\big)
+\begin{pmatrix} \hat{u}_{23} \\ \hat{v}_{23} \end{pmatrix} \big(\Theta_3 + \bar{\Theta}_3\big)\bigg],
\eea
where we define $z_2 = 2^\gm$ and $z_3 = (\sqrt{3})^\gm$,  and then 
\beas
\begin{pmatrix}
\hat{u}_{21} \\
\hat{v}_{21}
\end{pmatrix}= 
-\fl{1}{b^2}\begin{pmatrix}
0\\
1\end{pmatrix}; \qquad
\begin{pmatrix}
\hat{u}_{2j} \\
\hat{v}_{2j}
\end{pmatrix} = 
-\fl{1}{b^2r(z_j -1)^2}\begin{pmatrix}
-z_j b \\
z_jbr^2+r
\end{pmatrix}, \ \ j = 2, 3.
\eeas

At ${\mathcal O}(\veps^3)$, substituting (\ref{u1}) and (\ref{u2}) into (\ref{veps3}), we get
\bea\label{veps3-1}
{\mathcal L}\bu_3 = \p_\tau \bu_1 + {\bf c} \big(\Theta + \bar{\Theta}\big) \bigg(\fl{\hat{a}}{b}\hat{u}_1 + \sum_{1 \le j \le 3} \eta_j \big(\Theta_j + \bar{\Theta}_j\big)\bigg),
\eea
where the coefficient $\eta_j$ is computed by
\beas
\eta_j =\hat{u}_1^2 \hat{v}_1 +2 b \xi\big(\hat{v}_1\hat{u}_{2j} + \hat{u}_1\hat{v}_{2j} \big) + \xi\Big(\fl{1}{b} + br^2+2r\Big)\hat{u}_1\hat{u}_{2j}, \quad 1\le j \le 3.
\eeas
By simple calculation, we obtain that
\beas
\big(\Theta + \bar{\Theta}\big)\big(\Theta_1 + \bar{\Theta}_1\big) &=& 2\, |{\bf W}|^2\sum_{1 \le j \le 3} \big(W_j\,e_j+ \bar{W}_j\,\bar{e}_j\big),\\
\big(\Theta + \bar{\Theta}\big)\big(\Theta_2 + \bar{\Theta}_2\big) &=& \sum_{1 \le j \le 3} |W_j|^2\big(W_j\, e_j + \bar{W}_j\,\bar{e}_j\big) + {\mathcal T}_2 \\
\big(\Theta + \bar{\Theta}\big)\big(\Theta_3 + \bar{\Theta}_3\big) &=& 2 \sum_{1 \le j \le 3}\big(|{\bf W}|^2  - |W_j|^2\big)\big(W_j\,e_j+ \bar{W}_j\,\bar{e}_j\big) + {\mathcal T}_3, \qquad\ \   \eeas
where we have used the relation ${\bk}_1 + {\bk}_2 + {\bk}_3 = 0$. 
The terms ${\mathcal T}_l$ (for $l = 2, 3$) are the residual terms, which can be ignored in deriving the amplitude equations.  
Substituting the above results into (\ref{veps3-1}) and neglecting residual terms ${\mathcal T}_l$ yields 
\bea\label{veps3-2}
{\mathcal L}\bu_3=\p_\tau \bu_1 + {\bf c} \sum_{\substack{1 \le j, k, l \le 3 \\ j \neq k, l}}\Big(\big(2\eta_1 + \eta_2\big)|{W}_j|^2 + 2\big(\eta_1+\eta_3\big)\big(|W_l|^2 + |W_k|^2\big) + \fl{\hat{a}}{b}\hat{u}_1\Big)\big(W_j\,e_j+ \bar{W}_j\,\bar{e}_j\big).\
\eea
For a linear system ${\mathcal L}{\bf v} = {\bf r}$, the Fredholm solvability condition suggests  that the existence of a nontrivial solution is ensured if the right-hand vector ${\bf r}$ is orthogonal to the zero eigenvectors of the adjoint operator ${\mathcal L}^\star$.
Here, we have the zero eigenvector of  ${\mathcal L}^\star$ as: 
\bea\label{ustar}
u^\star = \Big(\fl{1+br}{br}, \ \ 1\Big)^T \bar{e}_j, \qquad j = 1, 2, 3. 
\eea
Combining (\ref{veps2}) and (\ref{veps3}) to obtain a system of  ${\mathcal L}\big(\veps^2\bu_2 + \veps^3\bu_3\big)$,  we then apply the Fredholm solvability condition to its right hand side  and get
\bea\label{amplitude-w}
\veps b(1-r^2)(br+1)\p_\tau W_j=2\xi \bar{W}_k\bar{W}_l+\veps\left[\big(2\eta_1 + \eta_2\big)|{W}_j|^2 + 2\big(\eta_1 + \eta_3\big)\big(|W_l|^2+|W_k|^2\big) + \hat{a}\right]W_j,\ 
\eea
for permutations of $j, k, l = 1, 2, 3$.

For notational simplicity, we denote $A_j(t):= A_{j, u}(t)$ for $j  = 1, 2, 3$.
Then equations (\ref{solution0623}), (\ref{expand-u}) and (\ref{u1}) indicate the relation: 
\beas
A_j(t) = \veps \hat{u}_1 W_j(\tau) + {\mathcal O}(\veps^2), \quad j = 1, 2, 3; \qquad  \tau = \veps^2 t; \qquad \veps^2\hat{a} = a - a_{\rm cr}.
\eeas
Here, we will only focus on the amplitude equations for $u$-component,  as $A_{j, v} = \hat{v}_1 A_{j, u}/\hat{u}_1$.
Substituting the above relation into (\ref{amplitude-w}) and reorganizing the terms yields the amplitude equations: 
\bea 
c_0\p_tA_j(t)= c_1(a-a_{\rm cr})A_j + c_2\bar{A}_k\bar{A}_l +  \Big(c_3 |A_j|^2+c_4\big(|A_k|^2+|A_l|^2\big)\Big)A_j, \quad
\label{amp1}
\eea
for the permutations of $j, k, l = 1, 2, 3$, where the coefficients:
\beas
c_0 = b^3(1-r^2)(br+1), \quad \ c_1 = b^2, \quad \  c_2 = 2b\xi, \quad \ c_3= 2\eta_1 + \eta_2, \quad \  c_4 = 2(\eta_1 + \eta_3).
\eeas
To study pattern selections, we will carry out the linear stability analysis on the amplitude equations (\ref{amp1}). 
Rewrite the amplitude function $A_j = \rho_j(t)\,e^{i \varphi_j(t)}$ (for $j = 1, 2, 3$). 
Substituting it into (\ref{amp1}) leads to the systems for density $\rho_j$ and phase $\varphi = \varphi_1 + \varphi_2 + \varphi_3$ as: 
\bea\label{density}
\begin{aligned}
c_0\p_t \rho_j(t) &= c_1(a-a_{\rm cr})\rho_j + c_2\rho_k \rho_l \cos(\varphi)+ \big(c_3\rho_j^2 + c_4(\rho_k^2 + \rho_l^2)\big)\rho_j, \\
c_0\p_t \varphi(t) &= -c_2\,\fl{\rho_1^2\rho_2^2 + \rho_1^2\rho_3^2 + \rho_2^2\rho_3^2}{\rho_1\rho_2\rho_3}\sin(\varphi),
\end{aligned}
\eea
for the permutation of $j, k, l = 1, 2, 3$. 
The density $\rho_1 = \rho_2 = \rho_3 = 0$ for a spatial homogeneous steady state, while for stripe patterns $\rho_1 \neq 0$ and $\rho_2 = \rho_3 = 0$. 
For the hexagon (or spot) patterns, the density $\rho_1 = \rho_2 = \rho_3 \neq 0$ and phase $\varphi = 0$ or $\pi$. 
Furthermore, the hexagon patterns with $\varphi = 0$ and $\pi$ are referred to as positive (denoted as $H_0$) and negative (denoted as $H_\pi$) hexagons, respectively. 
In the following, we will study the parameter regimes of stripe and spot patterns and their stability. 

\subsection{Stripe patterns }
\label{section3-1}

In the case of steady stripe patterns,  the density function $\rho_j$ reduces to
\bea\label{stripe}
\rho_{1}^s = \sqrt{-\frac{c_1(a-a_{\rm cr})}{c_3}}, \qquad \ \ \rho_{2}^s = \rho_{3}^s \equiv 0.
\eea
It implies that the stripe pattern exists when $c_3 < 0$, since $c_1 > 0$ and $a - a_{cr} > 0$.
To understand the stability of stripe patterns, we perform the linear stability analysis on the system 
(\ref{density}) around the steady state (\ref{stripe}).  
For  brevity, we omit the detailed calculations.
Here, we obtain the characteristic equation: 
\begin{equation*}
\big[c_1(a-a_{\rm cr}) +3 c_3 \big(\rho_{1}^{s}\big)^2 - c_0 \lambda\big]\big[\big(c_1(a-a_{\rm cr}) + c_4\big(\rho_{1}^{s}\big)^2- c_0 \lambda\big)^2-c_2^2 \big(\rho_{1}^{s}\big)^2\big]=0.
\end{equation*}
Substituting the value of $\rho_{1s}$ in (\ref{stripe}), we obtain: 
\begin{equation}\label{eig1}
\lambda_1= -\fl{2 c_1(a-a_{\rm cr})}{c_0}, \qquad 
\lambda_{2,3}= -\fl{c_1(a-a_{\rm cr})}{c_0c_3}\Big({c_4-c_3}\pm |c_2|\sqrt{\fl{c_3}{-c_1(a-a_{\rm cr})}}\Big).
\end{equation} 
It is evident that $\lambda_1$ is always negative.  
Note that $c_3 < 0$ to ensure the existence of stripes.  
To ensure the existence of steady stripes, we require that ${\rm Re}(\lambda_j) < 0$ for $j = 2, 3$,  which is true if the following conditions are satisfied: 
\beas
c_4 < c_3 < 0, \qquad c_1 > -\fl{c_3c_2^2}{(c_3 - c_4)^2(a-a_{\rm cr})}.
\eeas

\subsection{Hexagon patterns} 
\label{section3-2}

If steady hexagon (or spot) patterns exist,  their densities satisfy
\bea\label{spot}
\rho_{1} = \rho_{2} = \rho_{3} = \rho_h,
\eea
where  $\rho_h$ is defined implicitly by 
\bea\label{quad}
(c_3 +2 c_4)\rho_h^2 + |c_2| \rho_h + c_1\big(a-a_{\rm cr}\big) = 0.
\eea
It immediately implies that the positive density $\rho_h > 0$ exists only when $c_3+2c_4 < 0$. 
Then using the linear stability analysis and equation (\ref{quad}), we obtain the growth rate of perturbations as
\beas
\lambda_1 = -\fl{1}{c_0}\big[2c_1(a-a_{\rm cr})+|c_2|\rho_h\big], \qquad 
\lambda_{2, 3} = \fl{2}{c_0}\big[\big(c_3-c_4\big)\rho_h-|c_2|\big]\rho_h.
\eeas
 If $\rho_h > 0$ exists, there is always $\lambda_1 < 0$, and thus the spot pattens are stable if $(c_3 - c_4)\rho_h < |c_2|$. 
Summarizing the above discussion, we obtain the conditions for stable spot patterns as 
\beas
c_3 + 2c_4 < 0, \qquad (c_3 - c_4)\rho_h < |c_2|.
\eeas
Furthermore, if $c_2 > 0$ (resp. $ < 0$) the patterns are $H_0$ (resp. $H_\pi$) hexagons. 

\begin{figure}[htb!]
\centerline{\includegraphics[height=5.860cm,width=8.160cm]{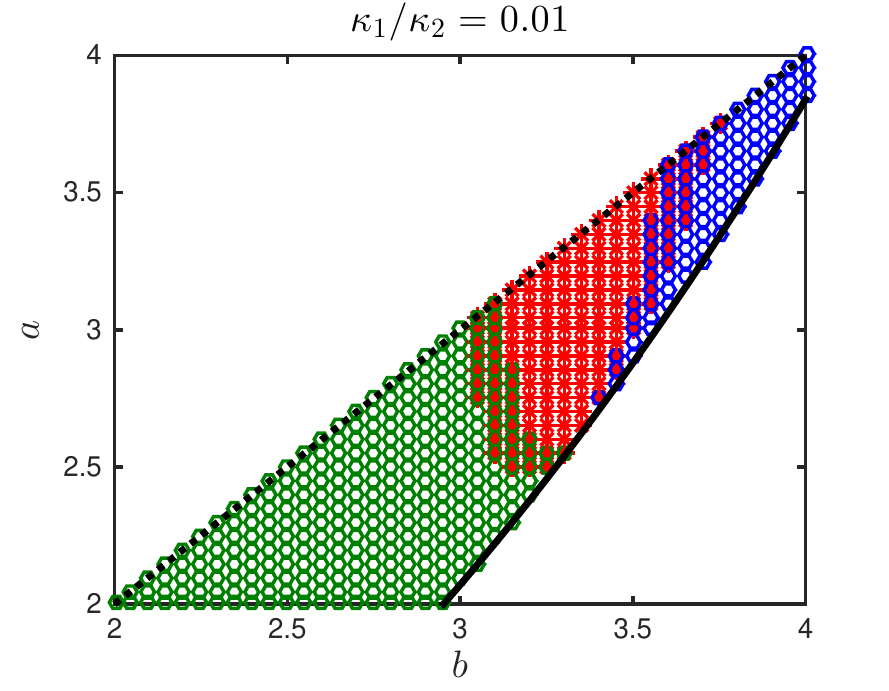}\hspace{-0cm}
\includegraphics[height=5.860cm,width=8.160cm]{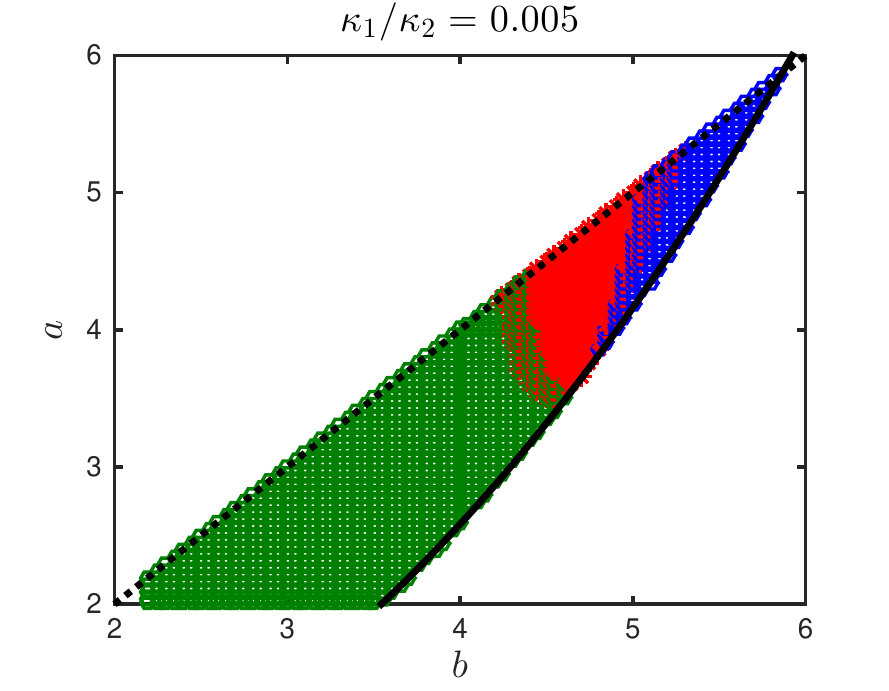}}
\caption{Illustration of stripe (red asterisk), $H_0$ hexagon (green circle), and $H_\pi$ hexagon (blue circle) in the Turing space for different diffusion ratios, where $\gamma_1 = \gamma_2 =2$.  }\label{Fig3-1}
\end{figure}

Figure \ref{Fig3-1} illustrates the parameter regimes of steady stripe and hexagon patterns in the Turing space of classical Schnakenberg equations. 
It shows that $H_0$ and $H_\pi$ hexagon patterns exist at the two ends of the Turing space, while stripe patterns are found between these two regions.
The overlap of stripe and spot regions are observed, where the stability conditions for both stripes and spots are satisfied. 
Hence, the mixed patterns of spots and stripes occur in the overlapping regions.
We find that even though the diffusion ratio affects the Turing space, the distribution regions of steady patterns are qualitatively the same. 
Additionally, the parameter regions for classical and fractional cases with $\gm_1 = \gm_2$ are almost the same, although their amplitude equations (\ref{amp1}) are different (because $c_3$ and $c_4$ depend on  $\gm$).

Our weakly nonlinear analysis predicts the parameter regimes for different patterns. 
In Section \ref{section4} and \ref{section5}, we will perform numerical simulations to study pattern formation in both classical and fractional Schnakenberg equations and compare them with our theoretical predictions. 
To this end, the two-dimensional Schnakenberg equation (\ref{SNG1}) with periodic boundary condition is discretized by the Fourier pseudospectral method in space and 4th order Runge--Kutta method in time. 
In our simulations, we will choose the  domain $\Og = [-4, 4]^2$ with number of grid points $N_x = N_y = 1024$ and time step $\Dt t = 0.005$. 
The initial condition is taken as the steady states $\bu_s$ with a small perturbation on $[-0.01, 0.01]^2$.
We have refined the mesh size and time step to make sure the conclusions are independent of these numerical parameters. 
In all pattern plots, only patterns of $u_1$ are presented, where red and blue represent the highest and lowest values of $u_1$, respectively. 
Unless otherwise stated, we will always choose $\kappa_2 = 1$ in the following simulations.

\section{Pattern formation with normal diffusion}
\label{section4}

So far, pattern formations in the  Schnakenberg equation have not been well understood, even in the classical (i.e., $\gamma_1 = \gamma_2 = 2$) cases.
Studies on some special parameters are reported in the literature \cite{Hao2020, Wu2016}, but no exhaustive report can be found on pattern formation and selection across different parameter regimes. 
To study the normal and anomalous diffusive effects,  we will thus start with patterns in the classical Schnakenberg equation with different  regimes of parameters $a$, $b$, and $r = \sqrt{\kappa_1/\kappa_2}$. 
Our extensive simulations show that patterns exist only when $a \ge a_{\rm cr}$, confirming the analytical results in Sec. \ref{section2}.
In particular, steady patterns at $a = a_{\rm cr}$ could vary greatly for different values of $b$.

\begin{figure}[htb!]
\centerline{\includegraphics[height=5.860cm,width=8.160cm]{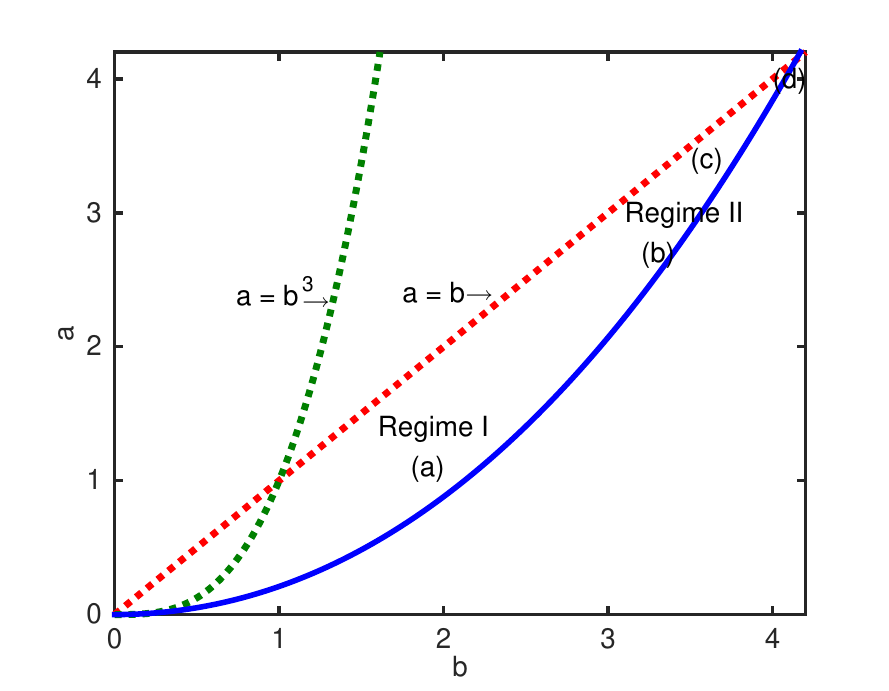}\hspace{-0cm}
\includegraphics[height=5.860cm,width=6.460cm]{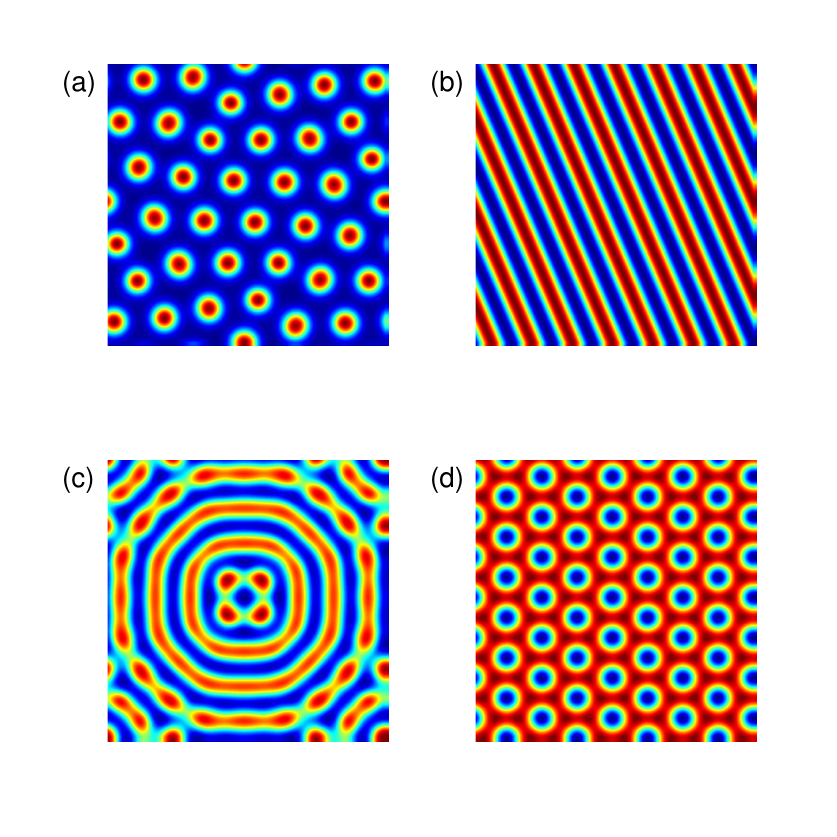}}
\caption{Illustration of  pattens in the classical Schnakenberg equation  with $\kappa_1 = 0.01$. }\label{Figure4-1}
\end{figure}

Figure \ref{Figure4-1} illustrates for representative patterns in the Turing space of the classical Schnakenberg equation with $\kappa_1 = 0.01$, where we could divide the Turing space into two regimes.  
In Regime I but $a \gg a_{\rm cr}$, only spot patterns are observed (i.e., pattern (a) in Fig. \ref{Figure4-1}).
By contrast, patterns in Regime II are more complicated. 
Various patterns,  including stripes, spots, and mixture of stripes and spots, are observed (see pattern (b)-(d) in Fig. \ref{Figure4-1}), depending on the combination of $a$, $b$ and $r$. 
Hexagon patterns are observed in both Regime I and II, which are $H_0$ hexagons in Regime I  and $H_\pi$ hexagons in Regime II (cf. patterns (a) and (d) in Fig. \ref{Figure4-1}).
These numerical observations confirm our weakly nonlinear analysis predictions in Fig. \ref{Fig3-1}.

\begin{figure}[htb!]
\centerline{
\includegraphics[height=3.260cm,width=3.560cm]{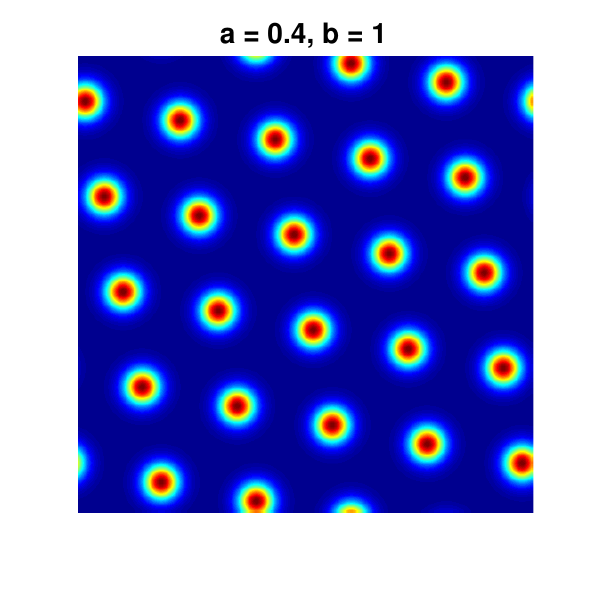}
\hspace{-6mm}
\includegraphics[height=3.260cm,width=3.560cm]{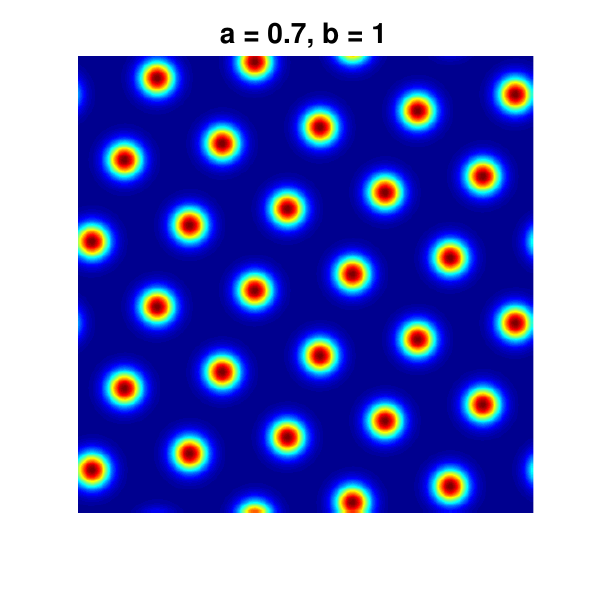}
\hspace{-6mm}
\includegraphics[height=3.260cm,width=3.560cm]{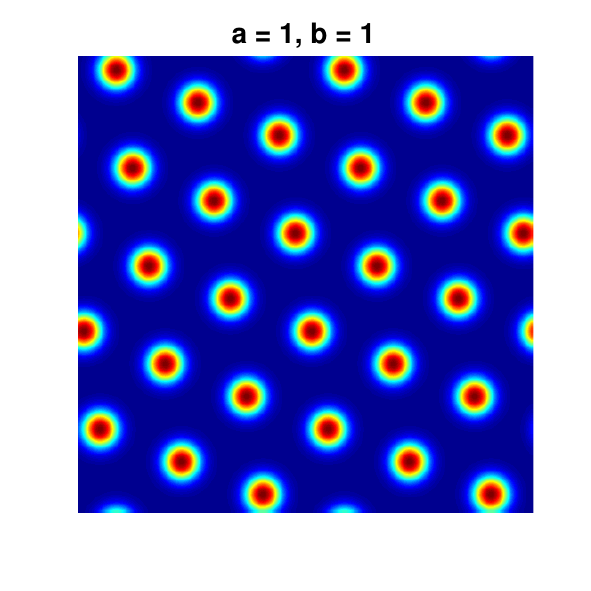}
\hspace{-5mm}
\includegraphics[height=3.260cm,width=4.560cm]{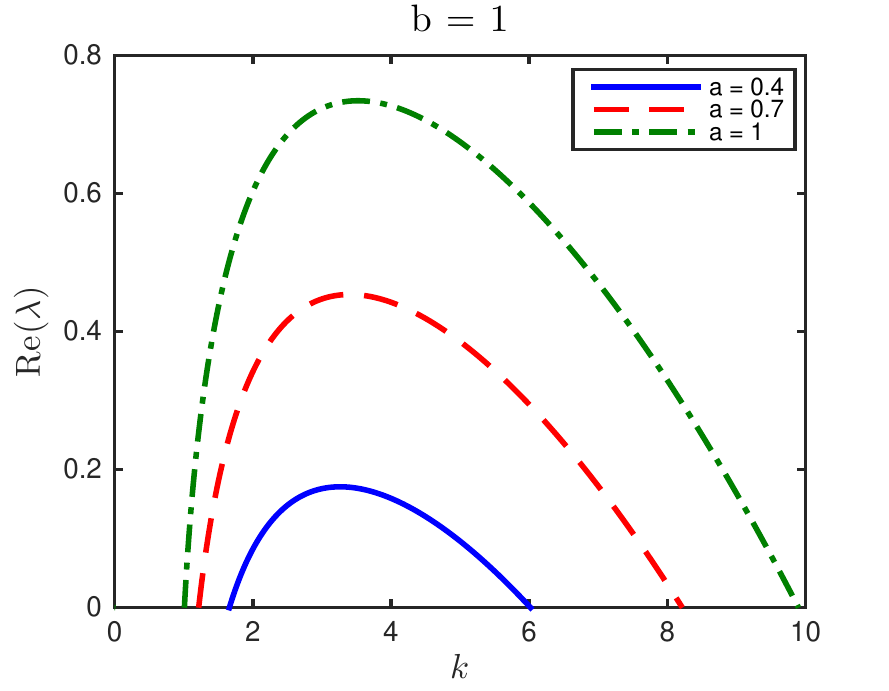}}
\centerline{
\includegraphics[height=3.260cm,width=3.560cm]{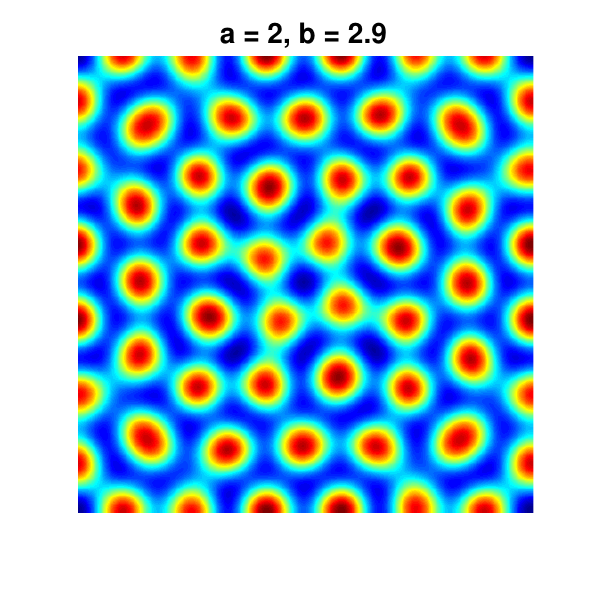}
\hspace{-6mm}
\includegraphics[height=3.260cm,width=3.560cm]{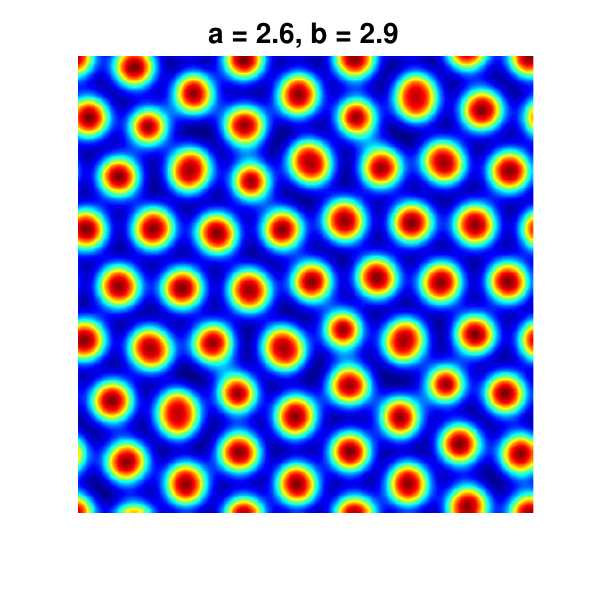}
\hspace{-6mm}
\includegraphics[height=3.260cm,width=3.560cm]{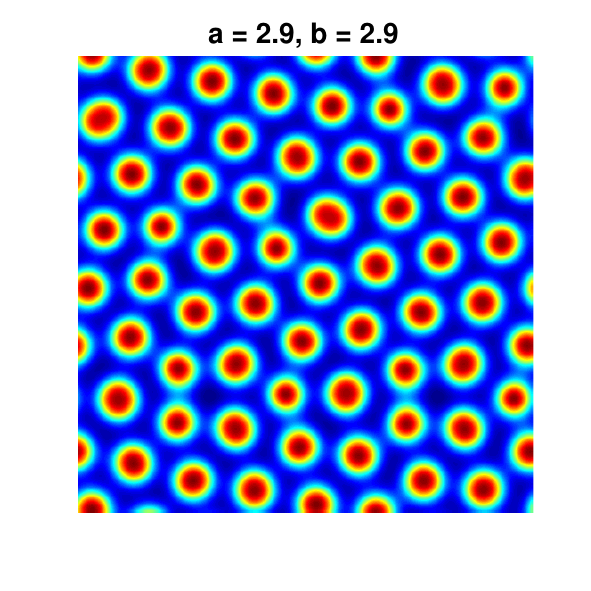}
\hspace{-5mm}
\includegraphics[height=3.260cm,width=4.560cm]{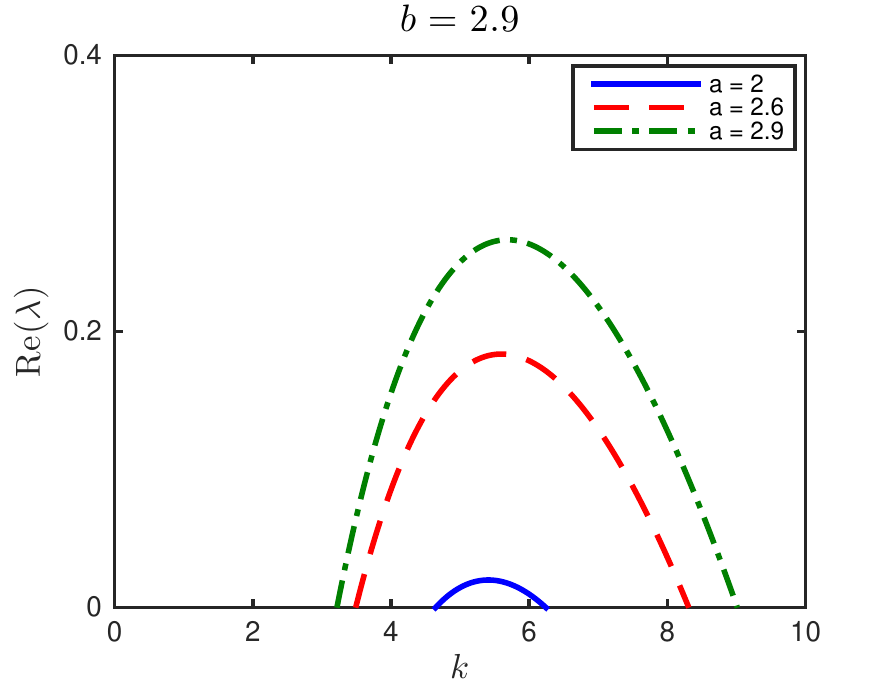}}
\centerline{
\includegraphics[height=3.260cm,width=3.560cm]{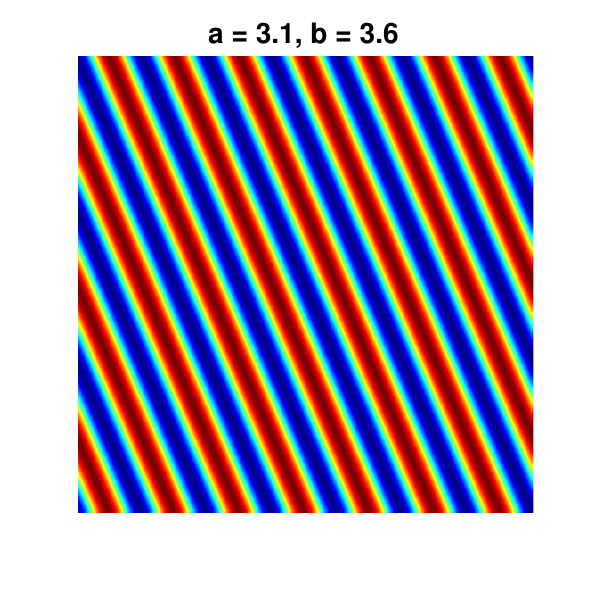}
\hspace{-6mm}
\includegraphics[height=3.260cm,width=3.560cm]{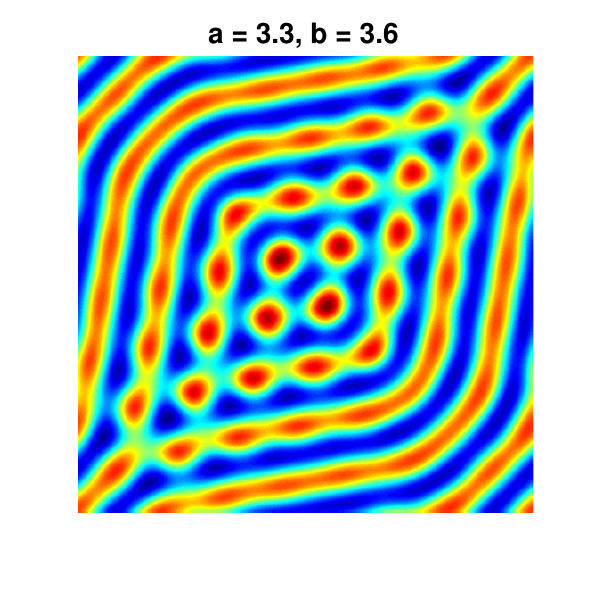}
\hspace{-6mm}
\includegraphics[height=3.260cm,width=3.560cm]{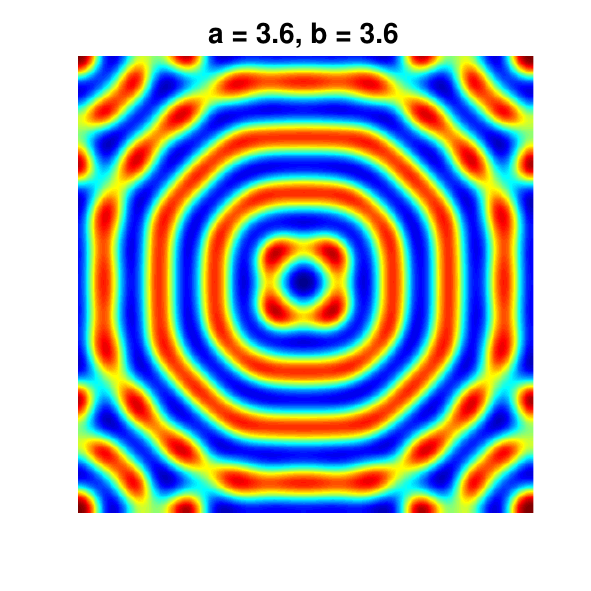}
\hspace{-5mm}
\includegraphics[height=3.260cm,width=4.560cm]{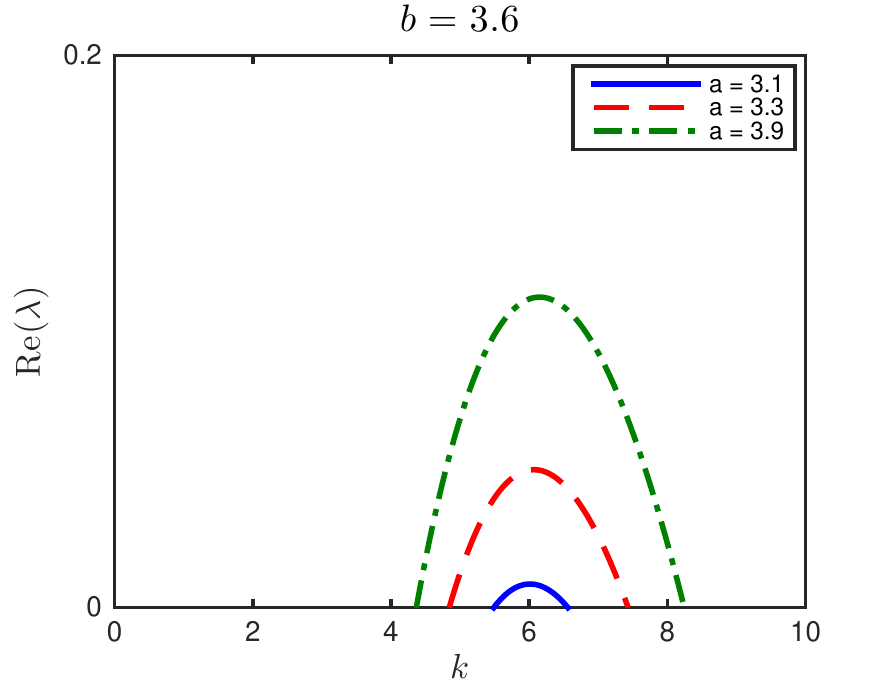}}
\caption{Patterns and dispersion relation in the classical Schnakenberg equation with  $\kappa_1 = 0.01$.} \label{Figure4-2}
\end{figure}

Figure \ref{Figure4-2} further demonstrates the patterns and corresponding dispersion relation for various $b$ and $a \in (a_{\rm cr}, b]$, where $\kappa_1 = 0.01$. 
For $b = 1$, the weakly nonlinear analysis shows that only spot patterns exist for any $a$, which is confirmed by our numerical results in Fig. \ref{Figure4-2}. 
It shows that the patterns are qualitatively the same, but the larger the value of $a$, the denser the spots.  
For different $a$, the dispersion relation reaches its maximum at the same wave number, but the growth rate and unstable band increase with $a$. 
We additionally find that in Regime I,  the density of spots in the steady patterns generally increases with the value of $b$.
Different from $b = 1$, spots start to connect when $b = 2.9$, locating in transition regime between I and II. 
The patterns for $b = 3.6$ are more complex, depending on the value of $a$.
Stripe patterns are observed for $a$ slightly larger than $a_{\rm cr} = 3.0586$ (see Fig. \ref{Figure4-2} with $a = 3.1$). 
As $a$ increases, the stripes start to deform,  and spots appear in the pattern, resulting in a mixed pattern of spots and stripes. 
It also shows that the growth rate in this case is much smaller. 
\begin{figure}[htb!]
\centerline{
\includegraphics[height=3.260cm,width=3.560cm]{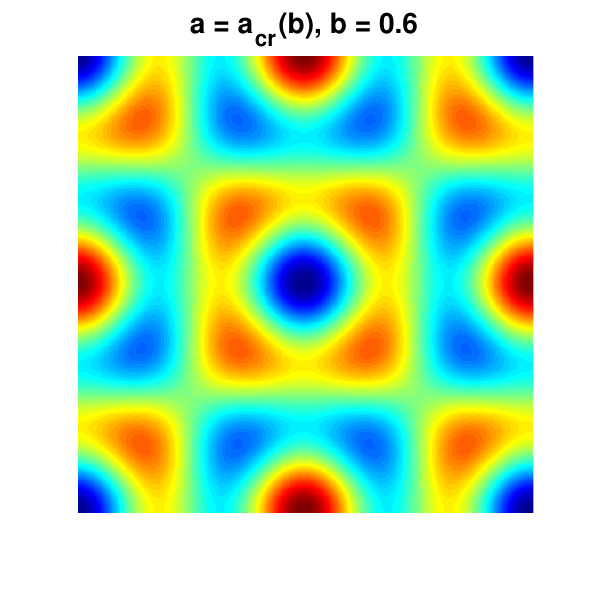} \hspace{-6mm}
\includegraphics[height=3.260cm,width=3.560cm]{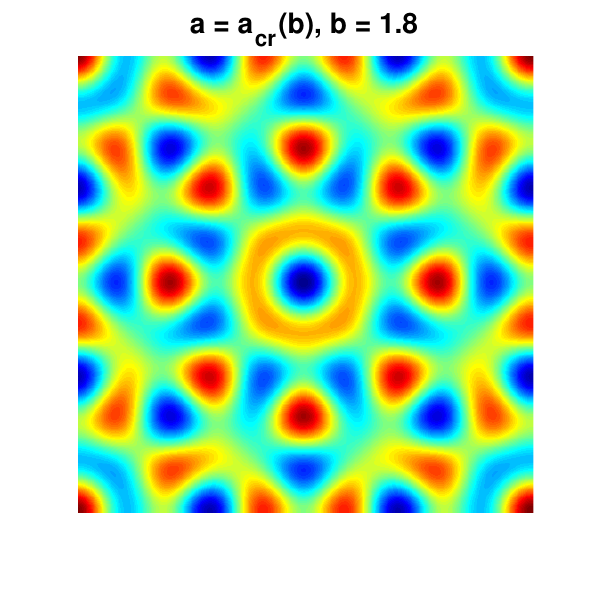} \hspace{-6mm}
\includegraphics[height=3.260cm,width=3.560cm]{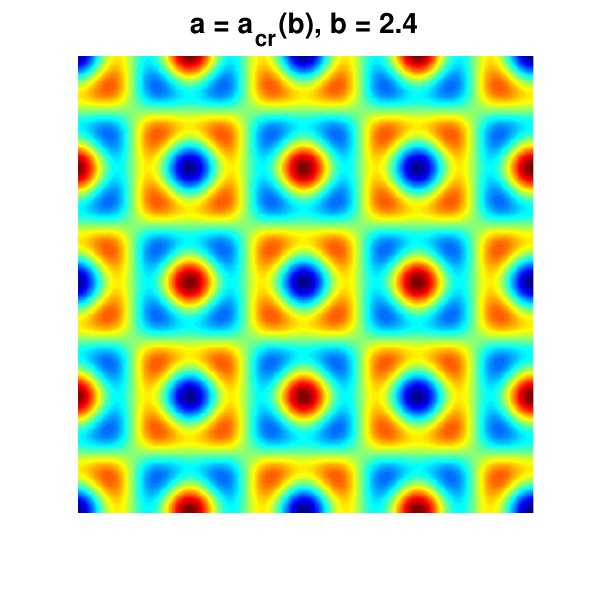} \hspace{-6mm}
\includegraphics[height=3.260cm,width=3.560cm]{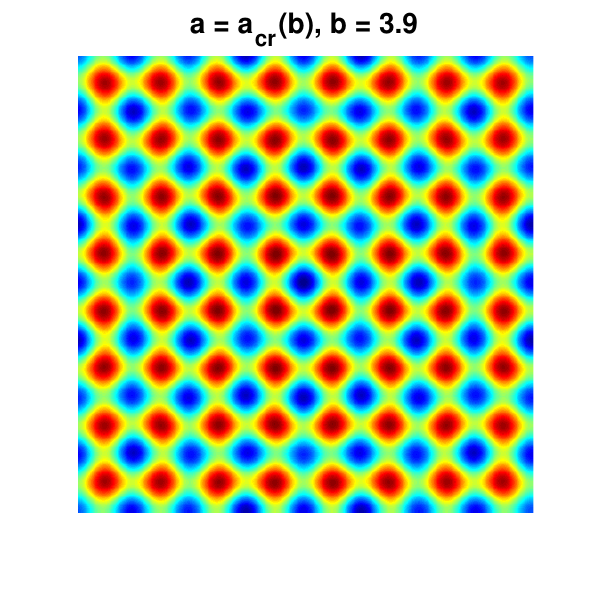} 
\hspace{-6mm}
\includegraphics[height=3.260cm,width=3.560cm]{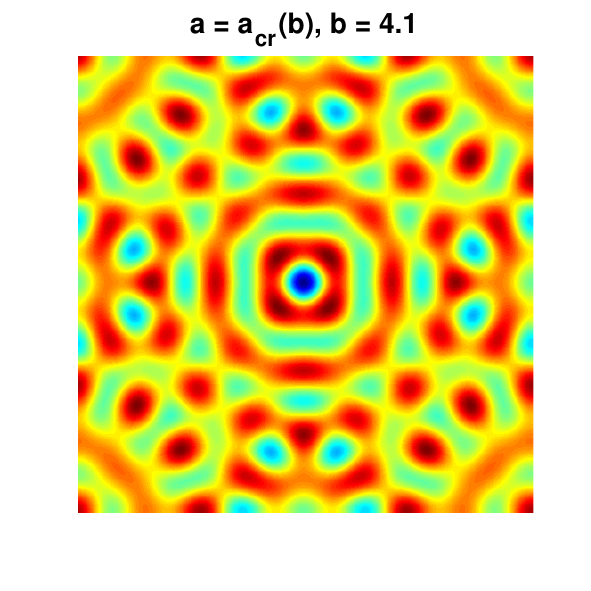}}
\caption{Patterns in the classical Schnakenberg equation with $\kappa_1 = 0.01$. }\label{Figure4-3}
\end{figure}

In Figure \ref{Figure4-3}, we study the patterns at the critical value, i.e., $a = a_{\rm cr}$. 
It shows that for a given $b$, patterns start appearing from $a = a_{\rm cr}$, but they are significantly different from those when $a \gg a_{\rm cr}$. 
Even though the analysis of amplitude equations predicts the existence of spot and stripe patterns,  it could not provide the information at the critical values. 
Hence, numerical studies play an important role in this regime. 
To the best of our knowledge, no reports of patterns at the critical value $a  = a_{\rm cr}$ can be found in the literature.

Next, we move to study the effects of diffusion coefficients $\kappa_1$ and $\kappa_2$ on pattern selection.   
Our linear stability analysis suggests that the Turing space depends only on the ratio $\kappa_1/\kappa_2$, and it expands as the  ratio decreases. 
For given parameters $a$ and $b$, the weakly nonlinear analysis further suggests that the steady patterns remain the same if ratio $\kappa_1/\kappa_2$ is the same.  
However, do the values of $\kappa_1$ and $\kappa_2$ play a role on in pattern formation? 
\begin{figure}[htb!]
\centerline{
\includegraphics[height=3.360cm,width=3.560cm]{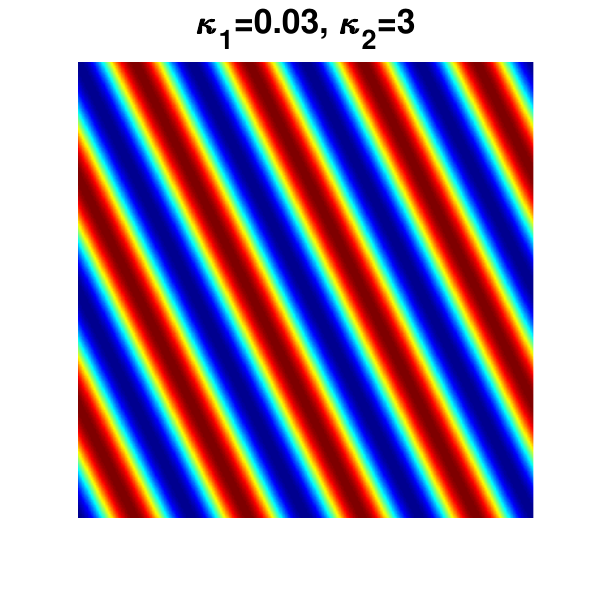}
\hspace{-6mm}
\includegraphics[height=3.360cm,width=3.560cm]{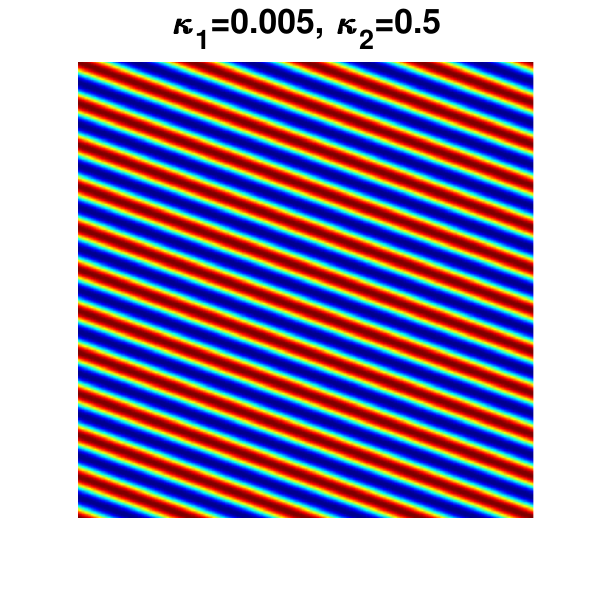}
\hspace{-6mm}
\includegraphics[height=3.360cm,width=3.560cm]{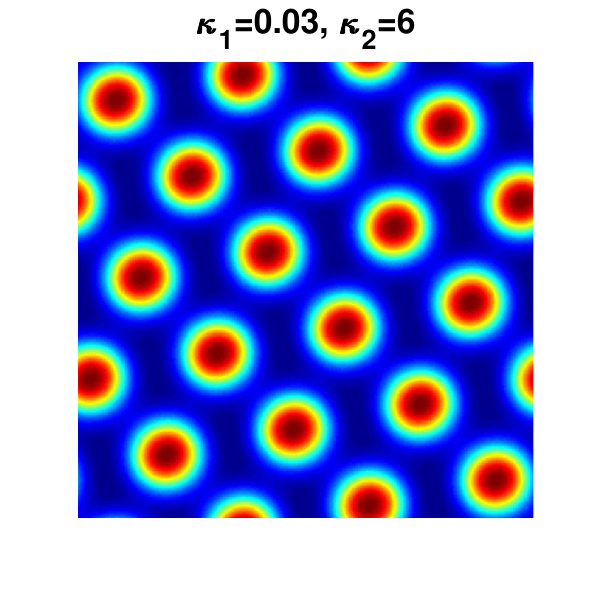}
\hspace{-6mm}
\includegraphics[height=3.360cm,width=3.560cm]{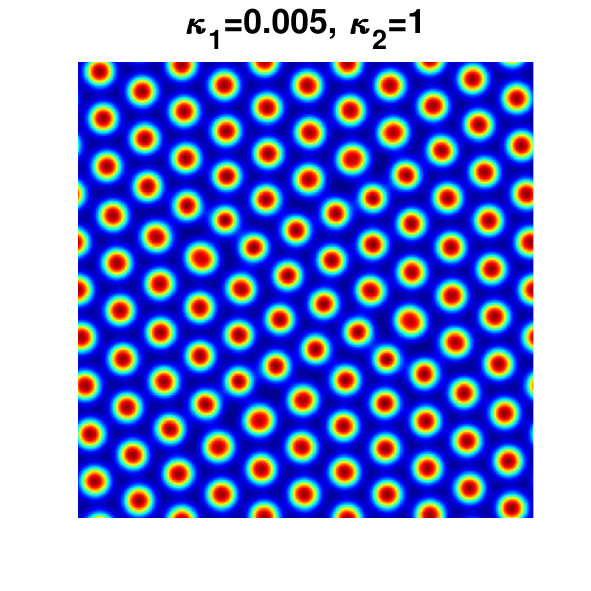}}
\caption{Patterns in the classical Schnakenberg equation with $a = 3.4$ and $b = 3.5$.} \label{Figure4-4}
\end{figure}
To understand it, we show the patterns for various $\kappa_1$ and $\kappa_2$ in Figure \ref{Figure4-4}, where  $b = 3.5$  and $a = 3.4$.
Note that the pattern for $\kappa_1 = 0.01$ and $\kappa_2 = 1$ can be found in Fig. \ref{Figure4-1} (c).
The same initial conditions and numerical parameters are used for Fig. \ref{Figure4-4} and Fig. \ref{Figure4-1} (c). 
For ratio $\kappa_1/\kappa_2 = 0.01$,  mixed and stripe patterns are observed in Fig. \ref{Figure4-1} (c) and Fig. \ref{Figure4-4}, respectively.  
This might be because $a = 3.4$ and $b = 3.5$ is on the boundary between the stripe region and mixed region (see Fig. \ref{Fig3-1}), and thus a small perturbation can change the steady patterns. 
As the ratio $\kappa_1/\kappa_2$ decreases, the $H_0$ spot patterns become more favorable, consistent with our prediction in Fig. \ref{Fig3-1}. 
On the other hand, the patterns generally remain the same for fixed $\kappa_1/\kappa_2$,  but their scales are much smaller with the decrease of product $\kappa_1\kappa_2$, which could be also understood from their dispersion relation in Fig. \ref{Figure4-5}.
\begin{figure}[htb!]
\centerline{
\includegraphics[height=4.560cm,width=6.560cm]{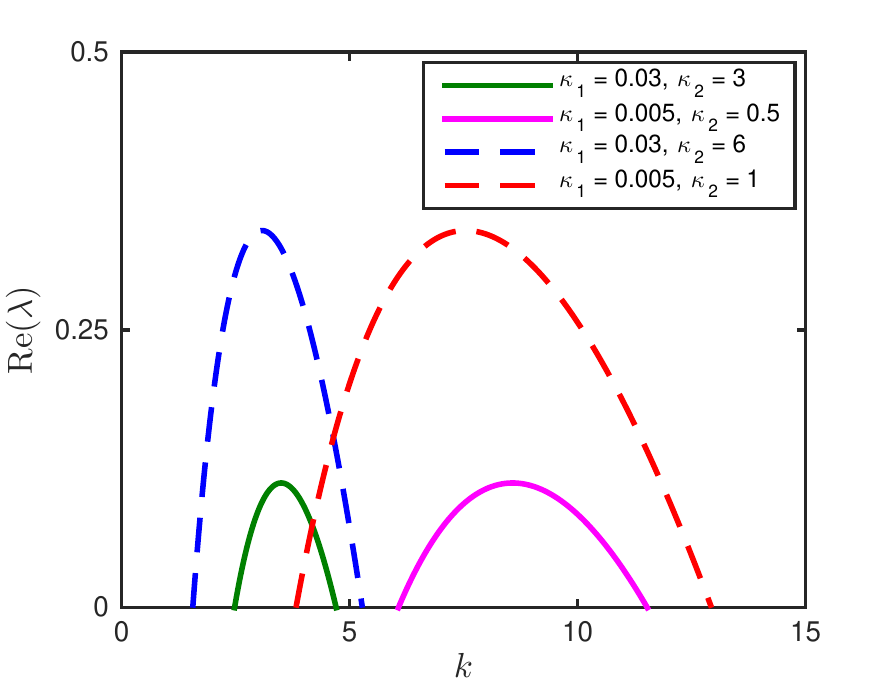}}
\caption{Dispersion relation for patterns in Figure \ref{Figure4-4}. }\label{Figure4-5}
\end{figure}
It shows that for the same diffusion ratio $\kappa_1/\kappa_2$, the smaller the product $\kappa_1 \kappa_2$, the larger the unstable wave numbers, and thus the finer the pattern scales. 
Hence,  the computations of patterns with smaller $\kappa_1\kappa_2$ become more challenging. 
The above observations  suggest the limitations of the linear stability analysis and weakly nonlinear analysis in the study of pattern formations.  

\section{Pattern formation  with superdiffusion}
\label{section5}

In this section, we study the pattern formation in the Schnakenberg equations when superdiffusion  is present in one or both components, i.e., $\gm_1, \gm_2 \le 2$.
We will divide our studies into two cases, i.e.,   $\gamma_1 = \gamma_2$,  and $\gamma_1 \neq \gamma_2$. 
The effects of superdiffusion on pattern formation will be studied by comparing to the results  in Section \ref{section4} for the classical Schnakenberg equations. 

\subsection{Same superdiffusion power $\gm_1 = \gm_2$}
\label{section5-1} 

For simplicity, we denote $\gm_1 = \gm_2 = \gamma$. 
Our linear stability analysis shows that the Turing space in this case is identical to that of the classical Schnakenberg equations. 
In other words, if $\gamma_1 = \gamma_2$, the diffusion powers play no role in the Turing instability, but  they may affect the pattern selections according to our weakly nonlinear analysis. 
\begin{figure}[htb!]
\centerline{
\includegraphics[height=3.260cm,width=3.5160cm]{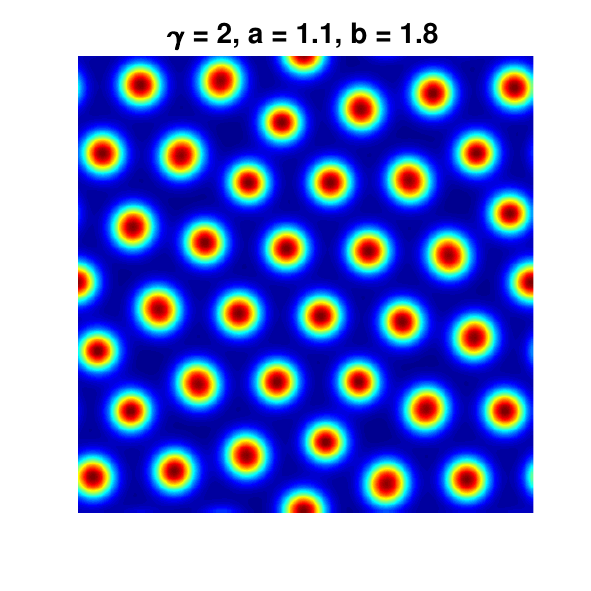}
\hspace{-6mm}
\includegraphics[height=3.260cm,width=3.5160cm]{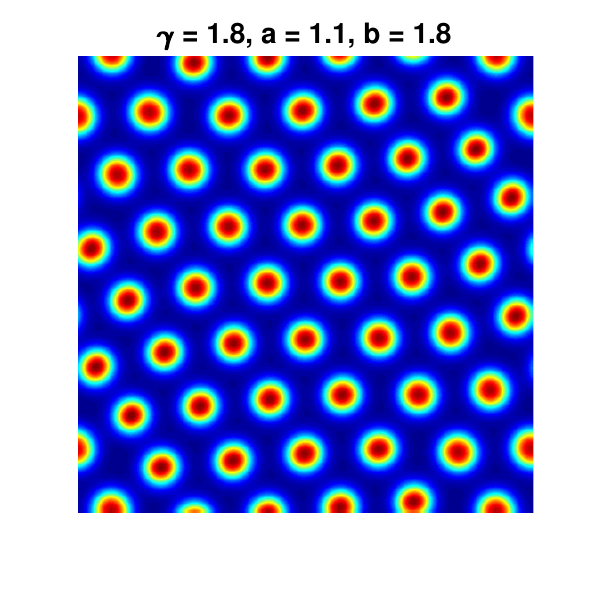}
\hspace{-6mm}
\includegraphics[height=3.260cm,width=3.5160cm]{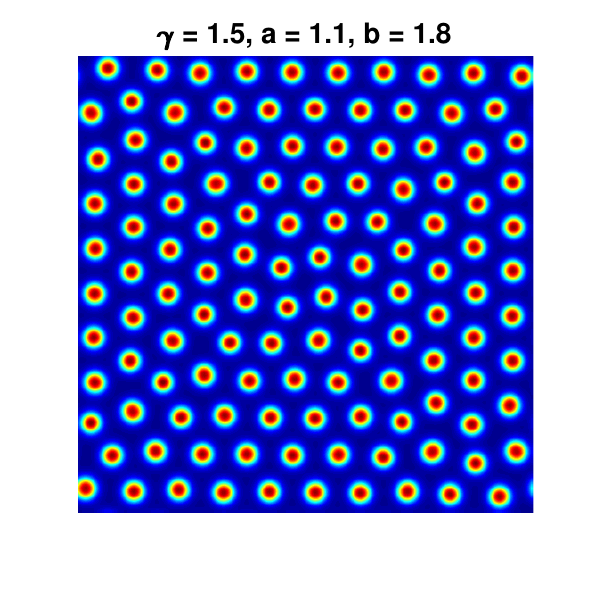}
\hspace{-6mm}
\includegraphics[height=3.260cm,width=3.5160cm]{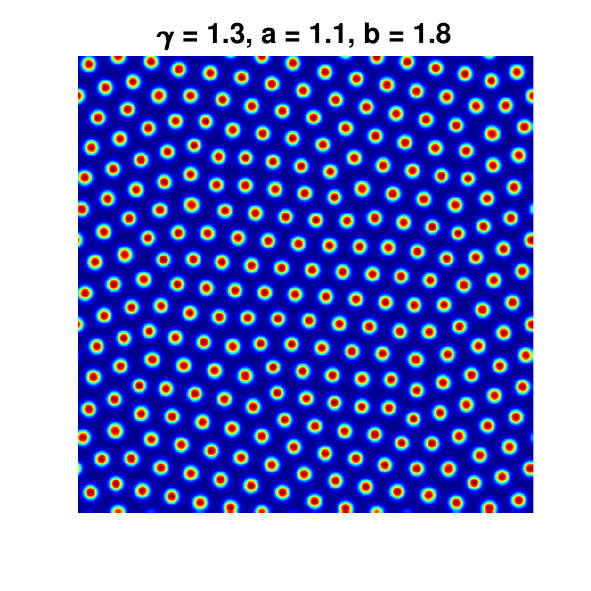}}
\vspace{-2mm}
\centerline{
\includegraphics[height=3.260cm,width=3.5160cm]{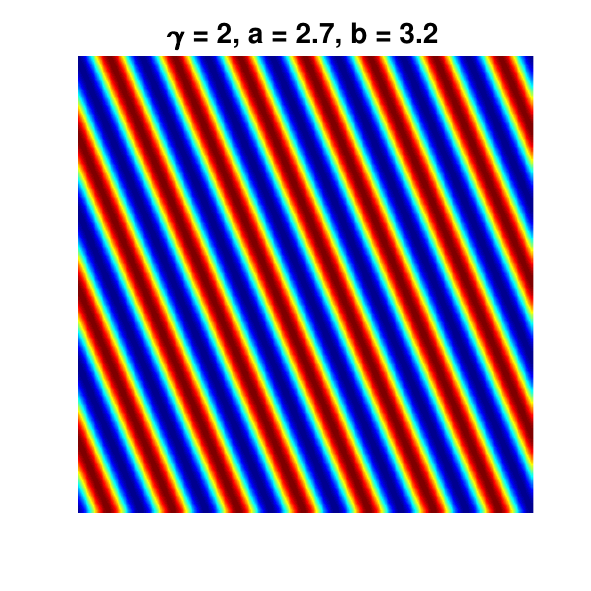}
\hspace{-6mm}
\includegraphics[height=3.260cm,width=3.5160cm]{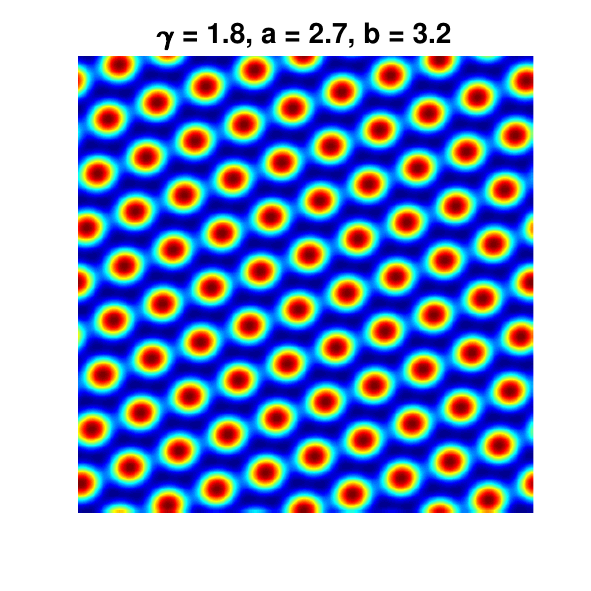}
\hspace{-6mm}
\includegraphics[height=3.260cm,width=3.5160cm]{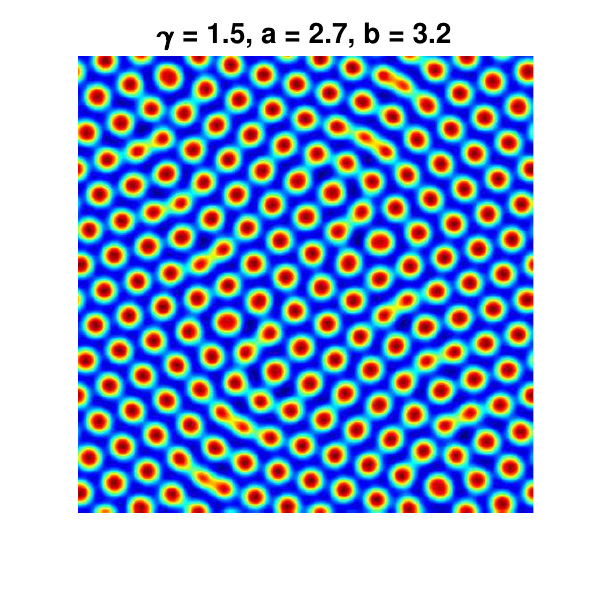}
\hspace{-6mm}
\includegraphics[height=3.260cm,width=3.5160cm]{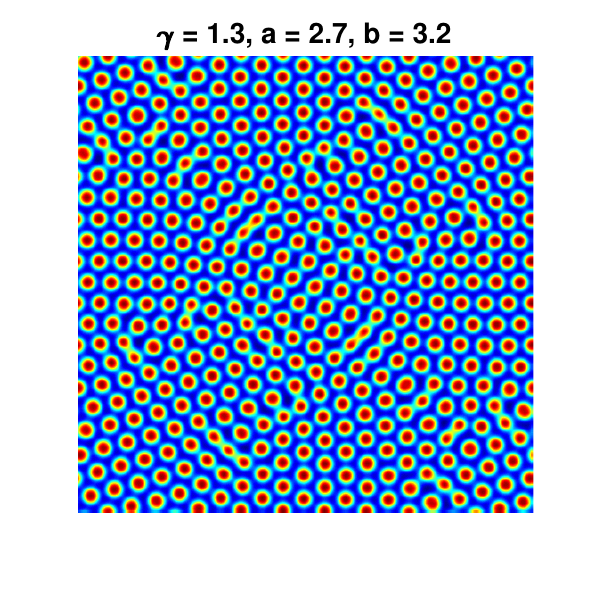}
}
\vspace{-2mm}
\centerline{
\includegraphics[height=3.260cm,width=3.5160cm]{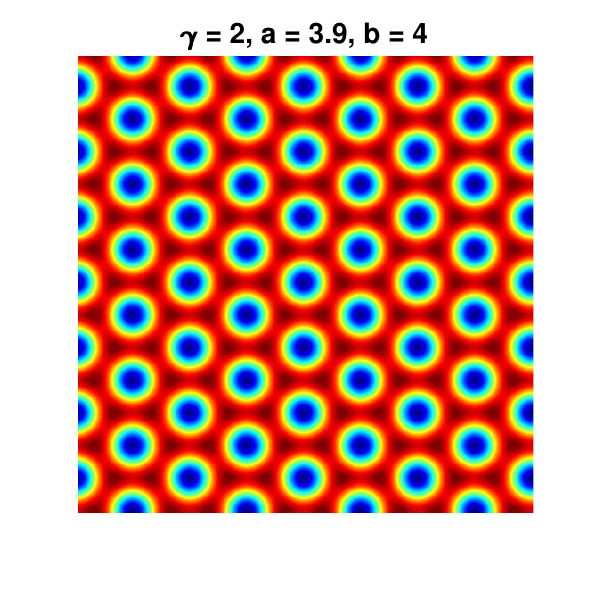}
\hspace{-6mm}
\includegraphics[height=3.260cm,width=3.5160cm]{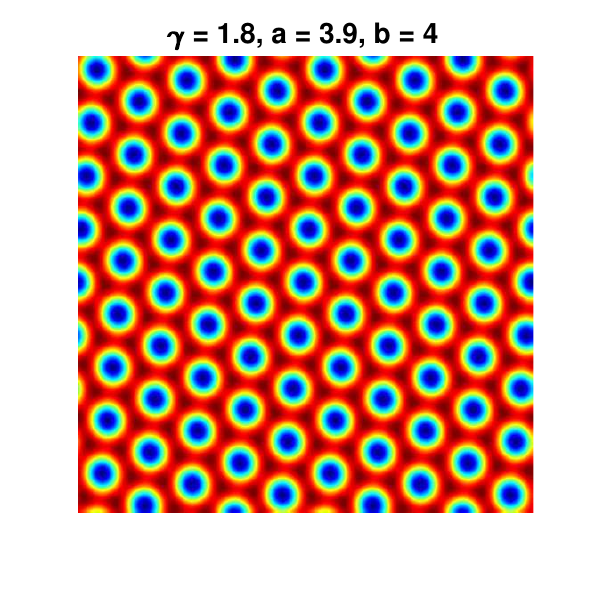}
\hspace{-6mm}
\includegraphics[height=3.260cm,width=3.5160cm]{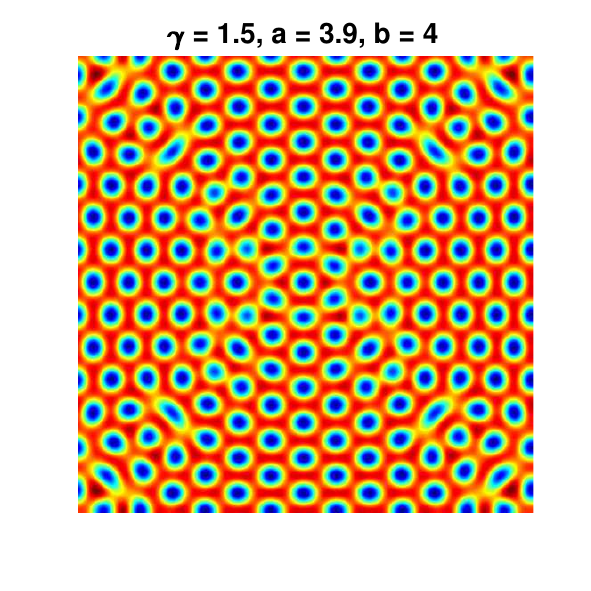}
\hspace{-6mm}
\includegraphics[height=3.260cm,width=3.5160cm]{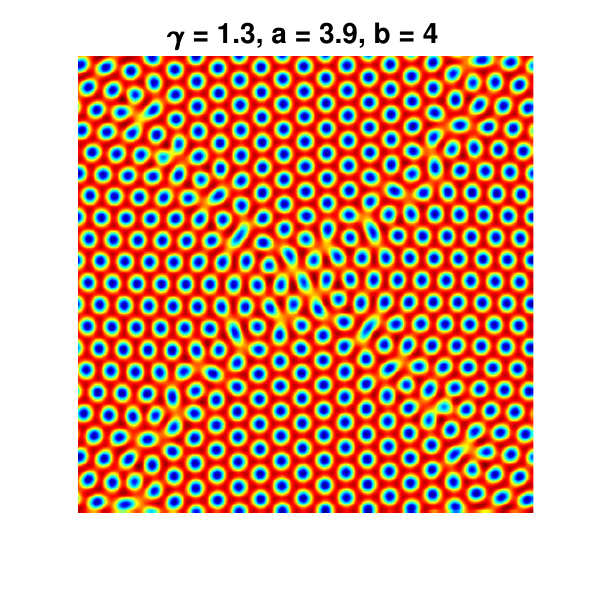}
}
\vspace{-2mm}
\centerline{
\includegraphics[height=3.260cm,width=3.5160cm]{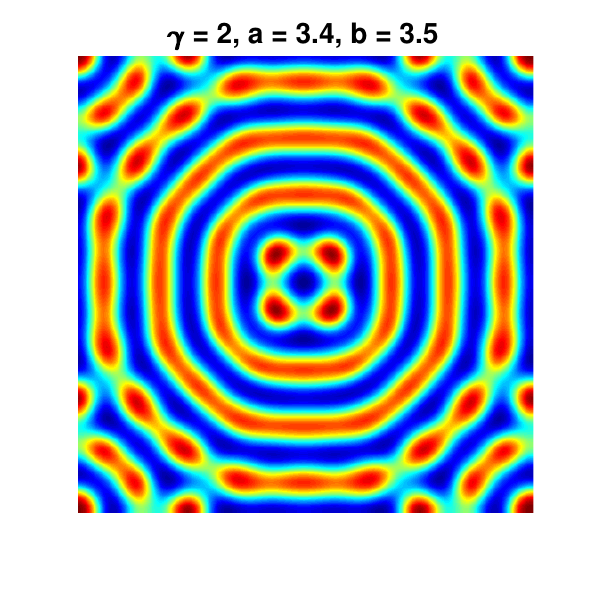}
\hspace{-6mm}
\includegraphics[height=3.260cm,width=3.5160cm]{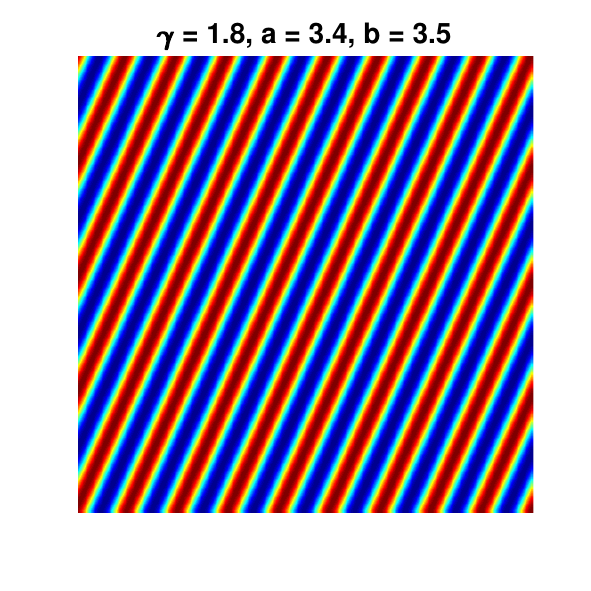}
\hspace{-6mm}
\includegraphics[height=3.260cm,width=3.5160cm]{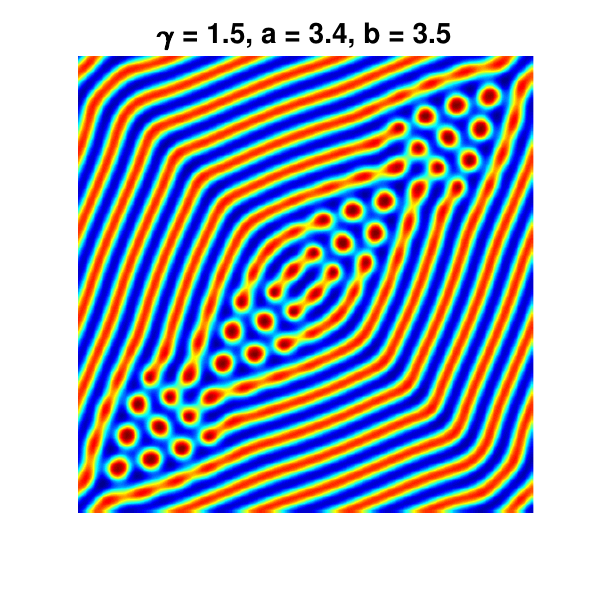}
\hspace{-6mm}
\includegraphics[height=3.260cm,width=3.5160cm]{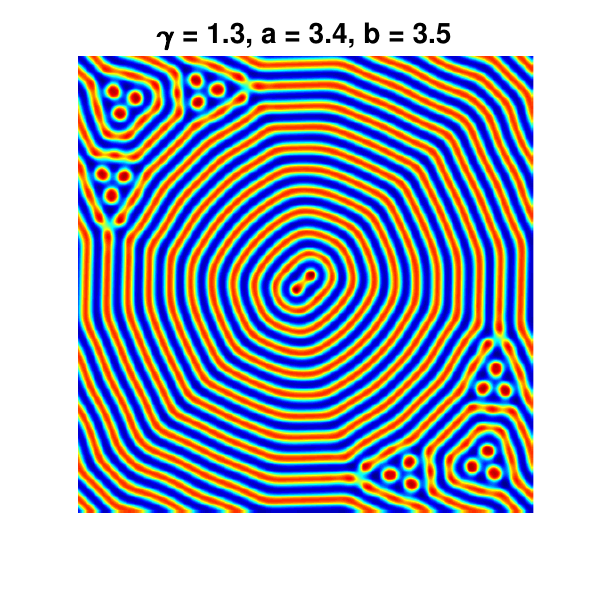}
}
\caption{Comparison of patterns in the classical and fractional Schnakenberg equations with $\kappa_1= 0.01$ and $\kappa_2 = 1$.} \label{Figure5-2}
\end{figure}
To further our understanding, we compare patterns in the classical and fractional cases in Figure \ref{Figure5-2}, and when $\gamma = 2$ they are the four representative patterns in Fig. \ref{Figure4-1}. 
It shows that the patterns in the classical and fractional Schnakenberg equations are qualitatively the same,  if $a$ and $b$ are far from the region of mixed patterns (see row 1 and 3 in Fig. \ref{Figure5-2}. 
However, if $a$ and $b$ are close to or in the region of mixed patterns, the superdiffusion has stronger effects on  pattern selection,  and a small change of power $\gamma$ could alter the type of patterns.
Generally,  the smaller the power $\gamma$, the stronger the superdiffusion, the finer the scales  of patterns. 
This can be also indicated in the dispersion relation in Fig. \ref{Fig2} (c) -- the smaller the power $\gamma$, the larger the unstable wave numbers, implying the finer scales of patterns. 
Computationally,  smaller mesh size and time step are demanded in order to capture the pattern details in the fractional cases, which greatly increases the computational costs and makes the simulations more challenging. 

\begin{figure}[htb!]
\centerline{
\includegraphics[height=3.260cm,width=3.560cm]{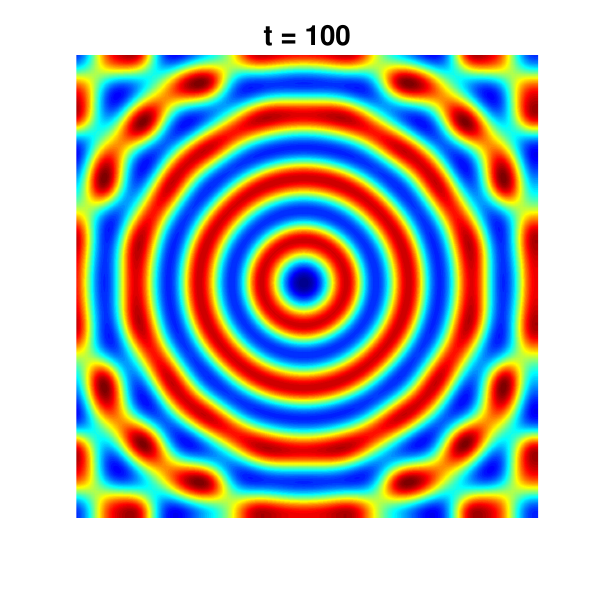}
\hspace{-6mm}
\includegraphics[height=3.260cm,width=3.560cm]{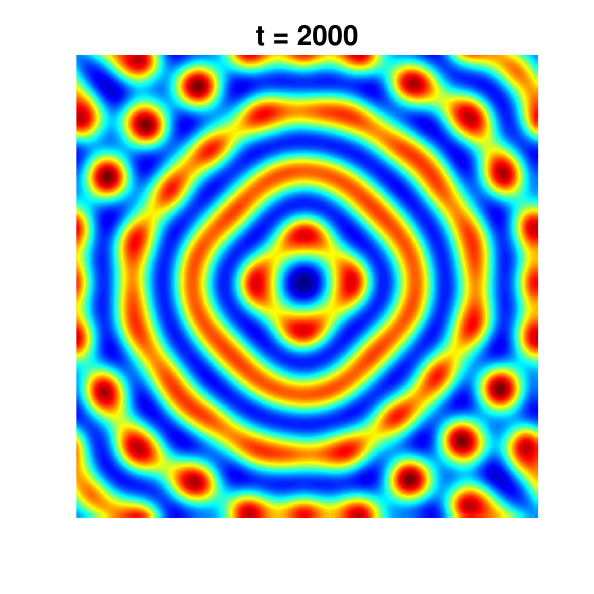}
\hspace{-6mm}
\includegraphics[height=3.260cm,width=3.560cm]{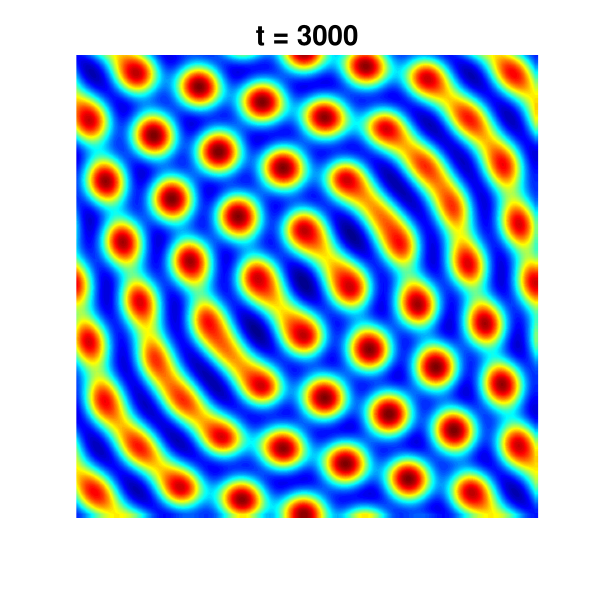}
\hspace{-6mm}
\includegraphics[height=3.260cm,width=3.560cm]{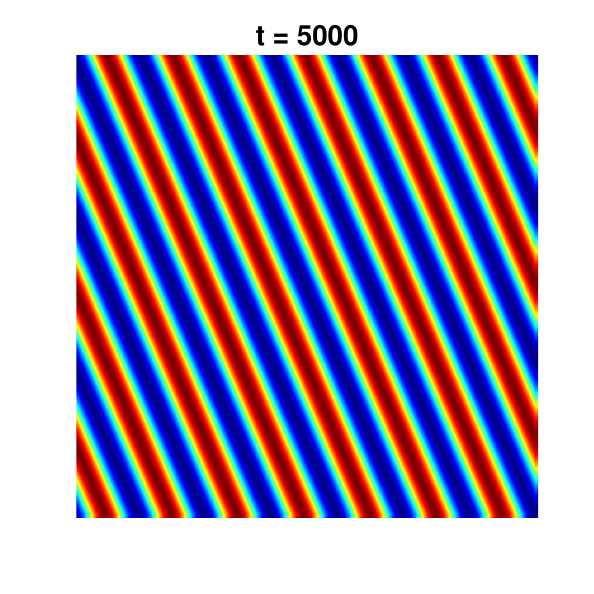}}
\centerline{
\includegraphics[height=3.260cm,width=3.560cm]{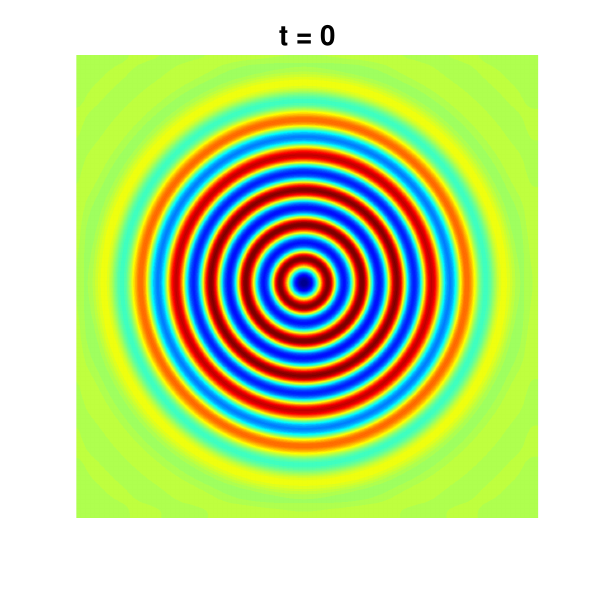}
\hspace{-6mm}
\includegraphics[height=3.260cm,width=3.560cm]{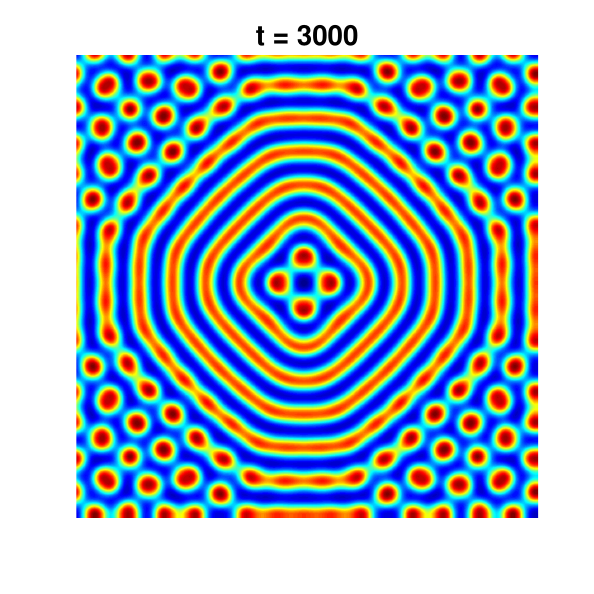}
\hspace{-6mm}
\includegraphics[height=3.260cm,width=3.560cm]{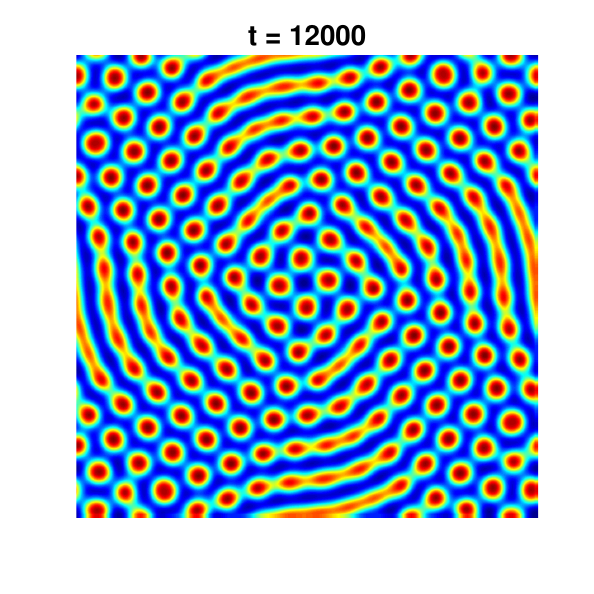}
\hspace{-6mm}
\includegraphics[height=3.260cm,width=3.560cm]{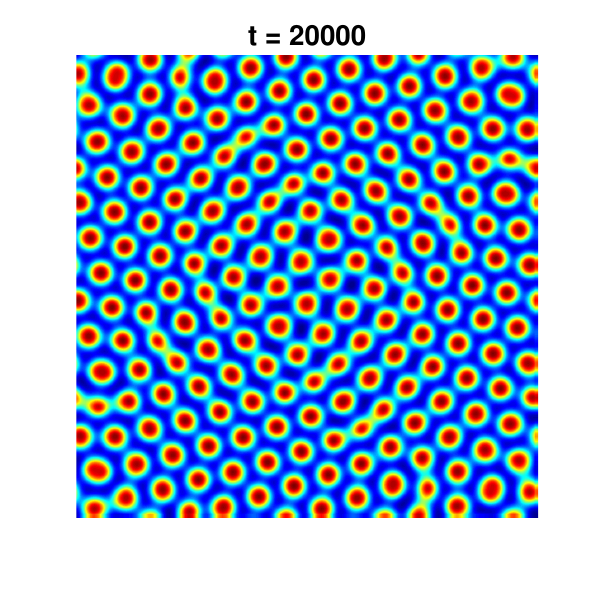}}
\caption{Dynamics of pattern formation in the classical (top row) and fractional (bottom row) Schnakenberg equations with $a = 2.7$ and $b = 3.2$.} \label{Figure5-3}
\end{figure}

In Figure \ref{Figure5-3}, we compare the time evolution of the pattern growth in the classical and fractional cases with $a = 2.7$ and $b = 3.2$.
The pattern initially emerges as spirals from the center of perturbation and then radiates towards the boundary. 
For both classical and fractional cases,  the spirals would break into spots once they reach the boundary. 
Then the spots in the classical cases will reconnect and form into steady stripe patterns, but remain spot patterns in the fractional cases. 
Moreover, we find that the fractional cases take much longer time to reach the steady patterns. 
We also study the effects of diffusion coefficients $\kappa_1$ and $\kappa_2$ on pattern formation and find similar results as in the classical cases. 
It shows that decreasing either the  ratio $\kappa_1/\kappa_2$ or power $\gm$ could lead to patterns with smaller scales, but the patterns from decreasing power $\gm$ are much denser. 
For brevity, we will omit showing these patterns here. 

\subsection{Different superdiffusion power $\gm_1 \neq \gm_2$}
\label{section5-2} 

\begin{figure}[htb!]
\centerline{
(a)\includegraphics[height=5.560cm,width=7.860cm]{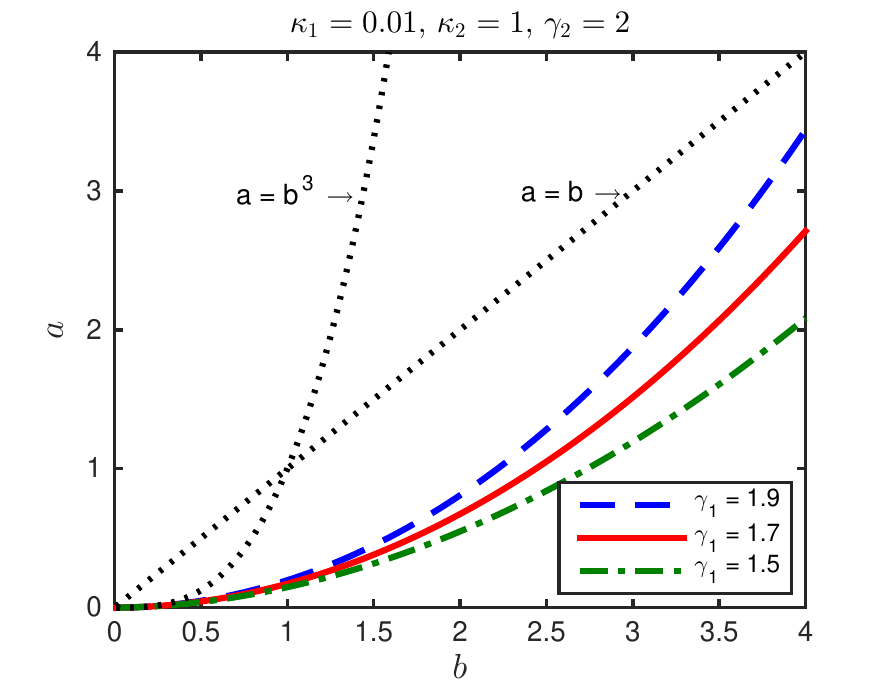}\hspace{-5mm}
(b)\includegraphics[height=5.560cm,width=7.860cm]{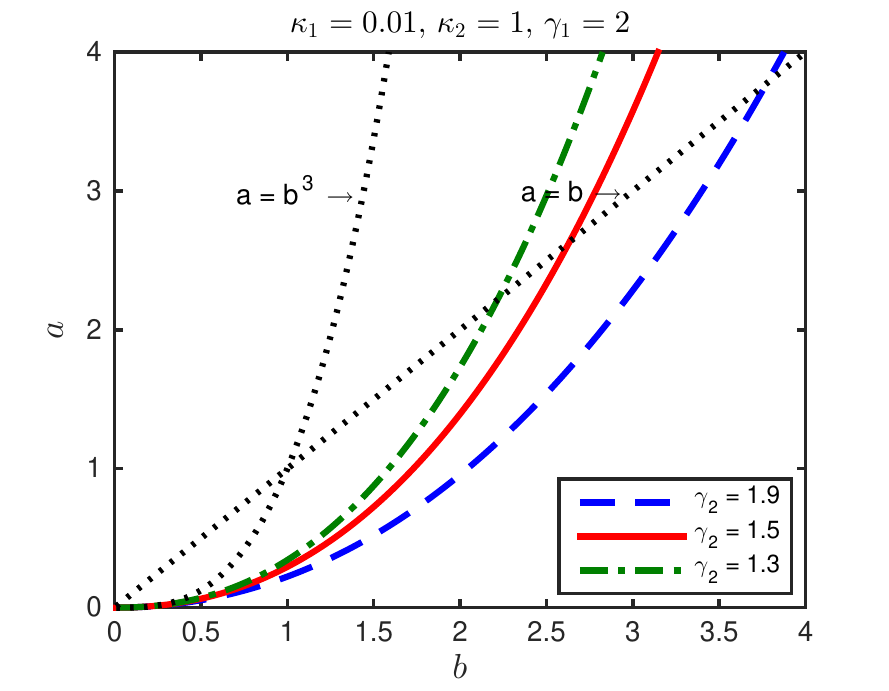}}
\caption{Comparison of Turing spaces (i.e., region between the non-dotted line and dotted-line of $\min\{b,\, b^3\}$) for different  diffusion powers $\gamma_1$ and $\gamma_2$. }
\label{Figure5-2-1}
\end{figure}

In the following, we explore the patterns in the  fractional Schnakenberg equations when two components have different diffusion powers. 
We will divide our discussion into two cases: $\gm_1 < \gm_2$ and $\gm_1 > \gm_2$.
Figure \ref{Figure5-2-1} compares the Turing spaces of different powers $\gamma_1$ and $\gamma_2$.
As discussed previously, the Turing space increases as the ratio $\gamma_1/\gamma_2$ decreases, and thus the spatially homogenous steady state $\bu_s$ is more unstable. 
The necessary conditions for the Turing instability is  $\kappa_1 < \kappa_2$, but the diffusion power $\gamma_1$ can be larger than $\gamma_2$.
Thus, the fractional models introduce more degrees of freedom to start patterns.

Figure \ref{Figure5-2-1} shows the patterns for $\gamma_1 < \gamma_2$, where we fix $\gamma_2 = 2$ and $\kappa_1 = 0.01$. 
It shows that even a small reduction of $\gamma_1$ leads to different patterns. 
For a fixed power $\gamma_2$,  decreasing  $\gamma_1$ would expand the Turing space and consequently enlarge the $H_0$ spot regions. 
\begin{figure}[htb!]
\centerline{
\includegraphics[height=3.260cm,width=3.560cm]{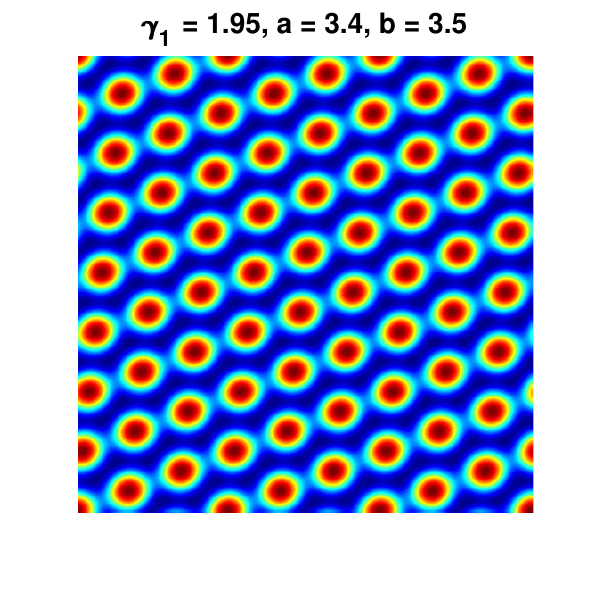}
\hspace{-6mm}
\includegraphics[height=3.260cm,width=3.560cm]{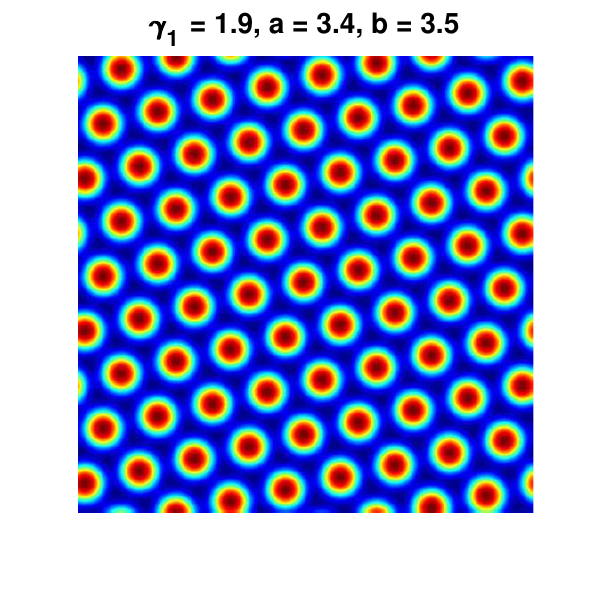}
\hspace{-6mm}
\includegraphics[height=3.260cm,width=3.560cm]{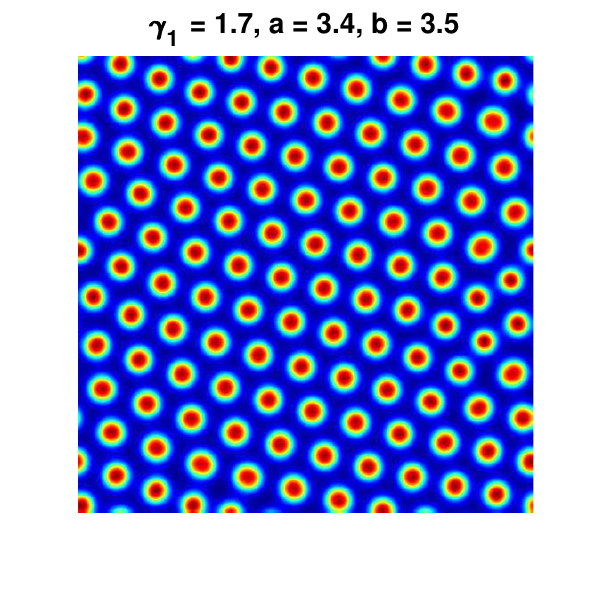}
\hspace{-6mm}
\includegraphics[height=3.260cm,width=3.560cm]{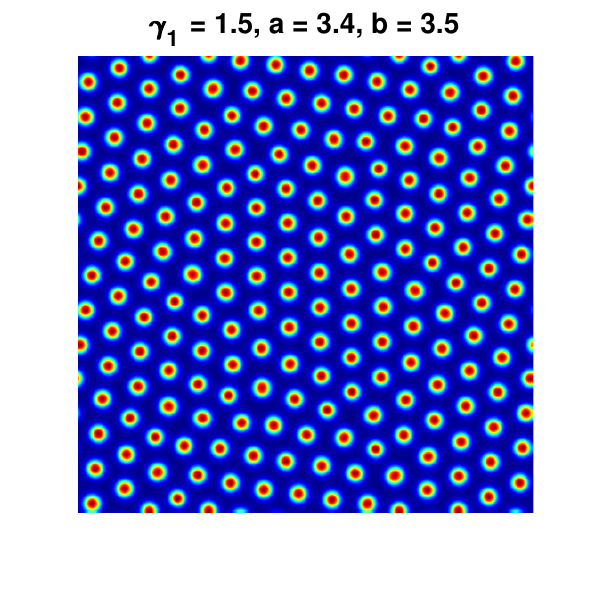}}
\vspace{-2mm}
\centerline{
\includegraphics[height=3.260cm,width=3.560cm]{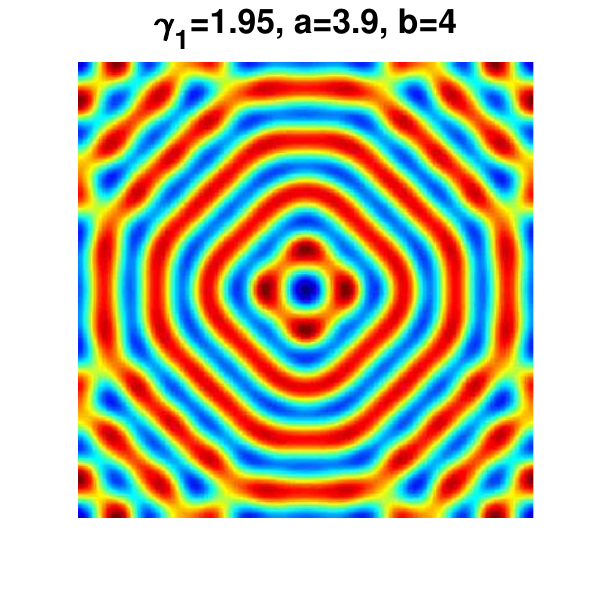}
\hspace{-6mm}
\includegraphics[height=3.260cm,width=3.560cm]{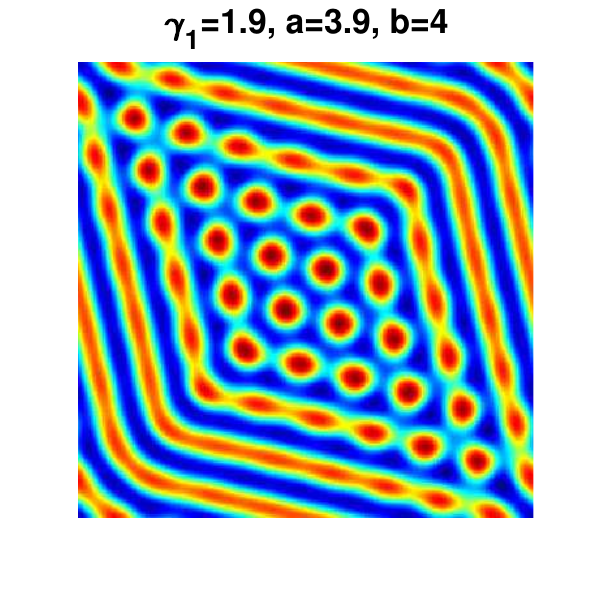}
\hspace{-6mm}
\includegraphics[height=3.260cm,width=3.560cm]{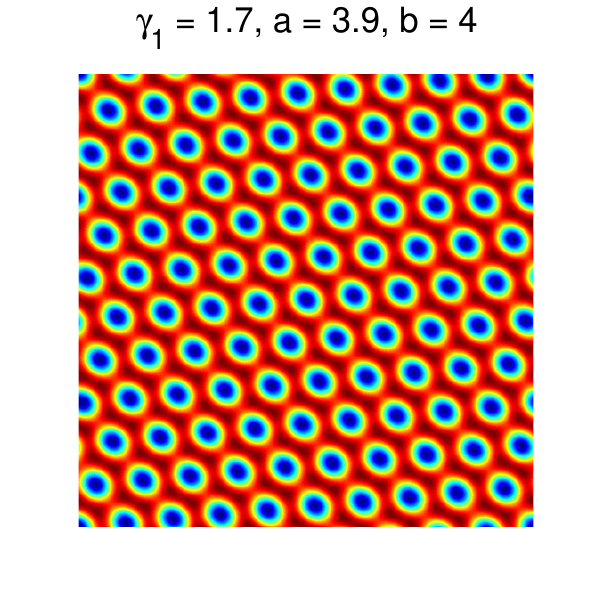}
\hspace{-6mm}
\includegraphics[height=3.260cm,width=3.560cm]{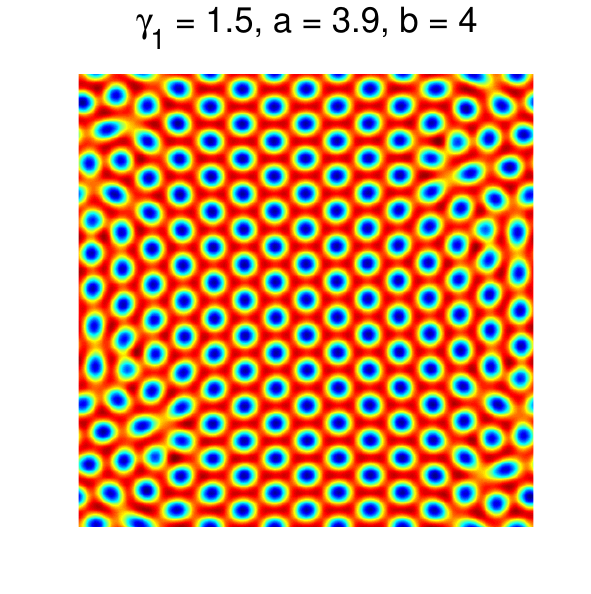}}
\caption{Patterns in the fractional Schnakenberg equations with $\gamma_2 = 2$, $\kappa_1 = 0.01$ and $\kappa_2 = 1$.}\label{Figure5-2-2}
\end{figure}
This is confirmed by our results in Fig. \ref{Figure5-2-2} -- the spot patterns become more favorable as $\gamma_1$ decreases. 
Moreover, the pattern scale reduces with the ratio $\gamma_1/\gamma_2$, as larger unstable wave numbers are presented (see Fig. \ref{Fig2} (b)).  

\begin{figure}[htb!]
\centerline{
\includegraphics[height=3.260cm,width=3.560cm]{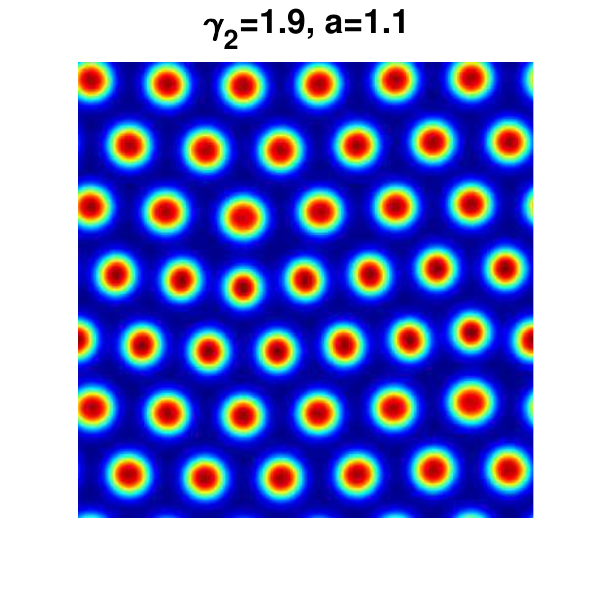}
\hspace{-6mm}
\includegraphics[height=3.260cm,width=3.560cm]{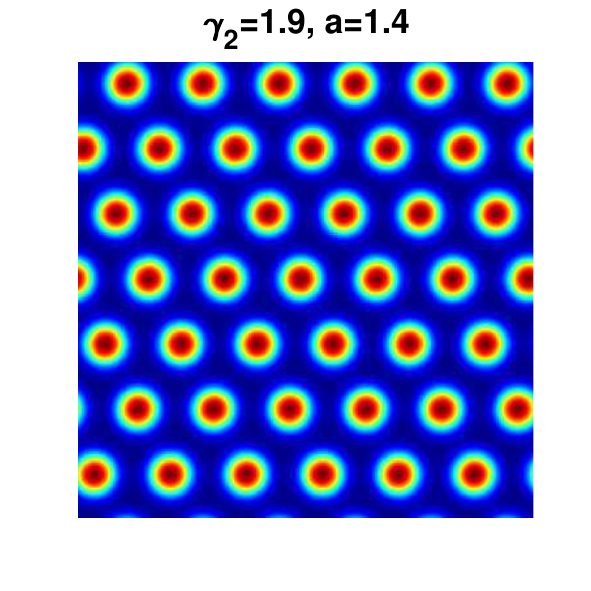}
\hspace{-6mm}
\includegraphics[height=3.260cm,width=3.560cm]{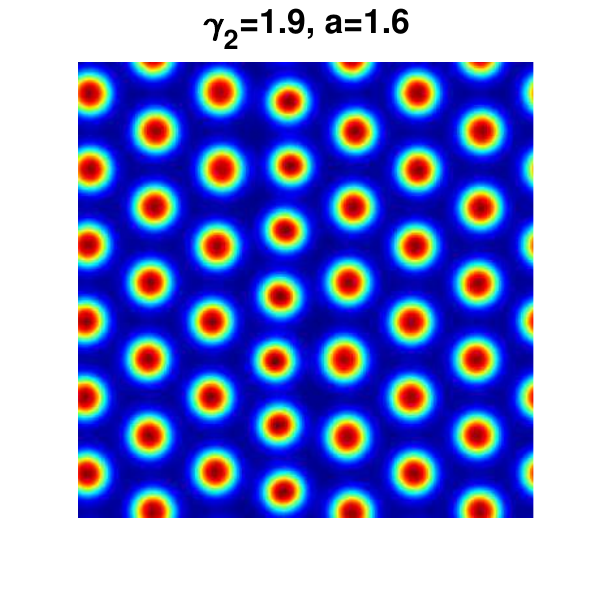}
\hspace{-6mm}
\includegraphics[height=3.260cm,width=3.560cm]{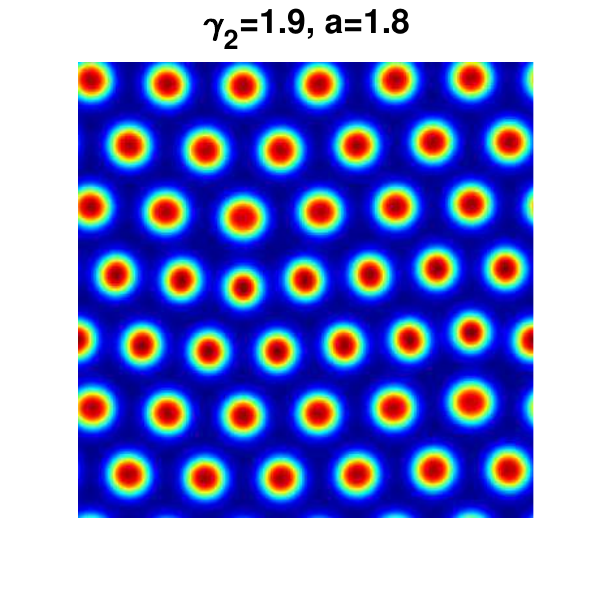}}
\centerline{
\includegraphics[height=3.260cm,width=3.560cm]{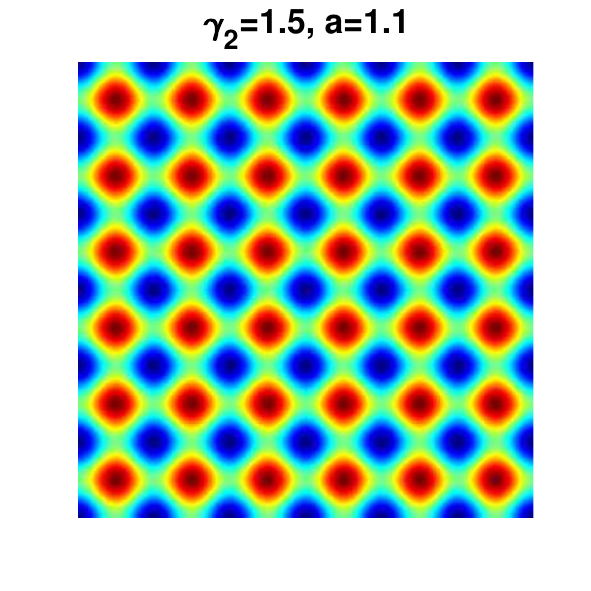}
\hspace{-6mm}
\includegraphics[height=3.260cm,width=3.560cm]{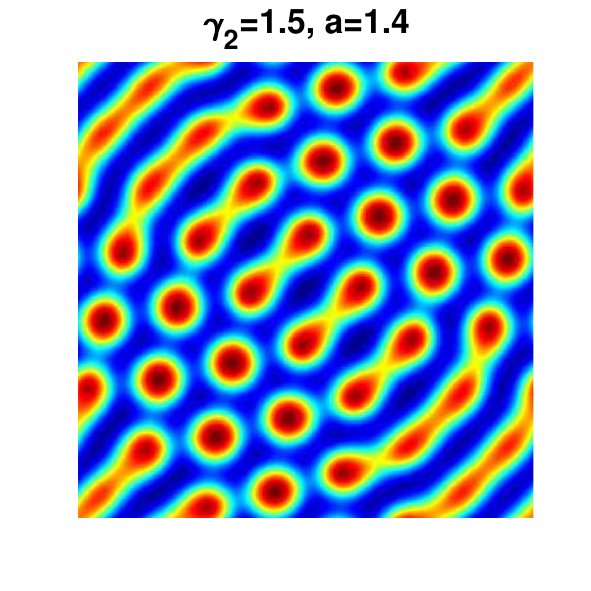}
\hspace{-6mm}
\includegraphics[height=3.260cm,width=3.560cm]{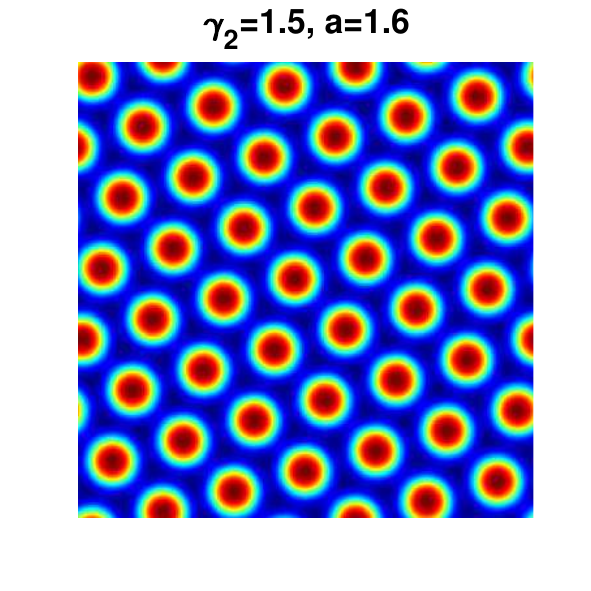}
\hspace{-6mm}
\includegraphics[height=3.260cm,width=3.560cm]{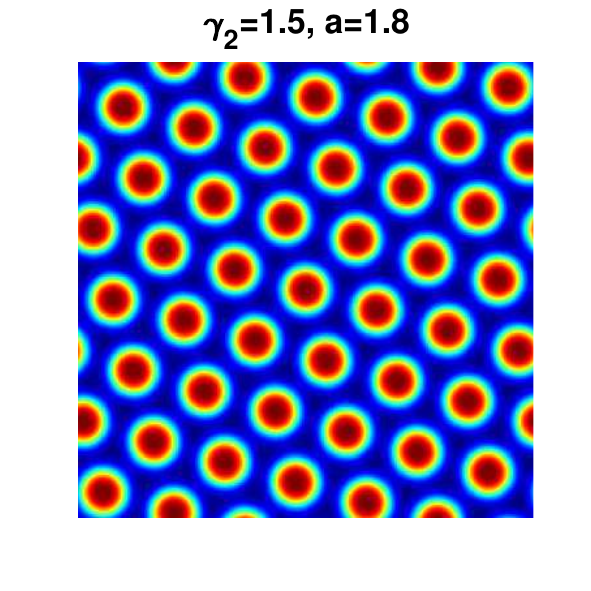}}
\caption{Patterns in the fractional Schnakenberg equation with $\gamma_1 = 2$, $\kappa_1 = 0.01$, $\kappa_2 = 1$, and $b = 1.8$.} \label{Figure5-2-3}
\end{figure}

On the other hand, Figure \ref{Figure5-2-3} presents the patterns for $\gamma_1 > \gamma_2$, where $\gamma_1  = 2$, $\kappa_1= 0.01$, and $b = 1.8$ are fixed.
For $\gamma_2 = 1.9$, the patterns are located in the $H_0$ spot regime, and thus similar patterns are observed for all $a_{\rm cr} < a < b$. 
With the decrease of $\gamma_2$, the Turing space quickly shrinks (see Fig. \ref{Figure5-2-1} (b)), and the parameter $b = 1.8$ is now around the boundary of $H_0$ spots and mixed pattern regions. 
Thus different patterns may be observed for different values of $a$ (see Fig. \ref{Figure5-2-3} with $\gamma_2 = 1.5$).
Note that the dispersion relation of this case is similar to that in Fig. \ref{Fig2} (a).

\section{Conclusions}
\label{section6}

We studied the pattern formation in the classical and fractional Schnakenberg equations and compare the effects of normal and super diffusion on pattern selection. 
Our studies not only provide a systematic understanding of Turing patterns in the classical Schnakenberg equations but also present detailed comparisons of patterns in classical and fractional models with different parameters. 
Our linear stability analysis suggested that the Turing space depends on both ratios of $\kappa_1/\kappa_2$ and $\gamma_1/\gamma_2$, which implies that the classical and  fractional model with $\gamma_1 = \gamma_2$ have the same Turing space. 
The Turing space increases as the ratio $\kappa_1/\kappa_2$ or $\gamma_1/\gamma_2$ decreases, and the necessary condition of Turing instability is $\kappa_1 < \kappa_2$. 
On the other hand, the unstable wave numbers and their growth rates  are sensitive to the values of $\kappa_l$ and $\gamma_l$ for $l = 1, 2$. 
Generally, the smaller the power  $\gamma_1$, the larger the unstable wave numbers and the smaller the pattern scales.  
Our weakly nonlinear analysis predicted the parameter regimes where hexagons, stripes, and their coexistence are expected in the Turing space. 
We numerically explored the interactions of diffusion coefficients and diffusion powers on the emergence of Turing patterns. 
Our numerical results confirmed the theoretical analysis and also provided new insights on the patterns in the fractional Schnakenberg equations. 

\bb
{\bf Acknowledgements.\ }   This work was supported by the US National Science Foundation under Grant Number DMS-1620465.


\end{document}